\documentclass[twocolumn,prl,floatfix,superscriptaddress,nobibnotes]{revtex4}

\usepackage{verbatim}
\usepackage{amsmath}
\usepackage{amssymb}
\usepackage{graphicx}
\usepackage{hyperref}
\usepackage{multirow}
\usepackage{array}
\usepackage{float}
\usepackage{enumitem}
\usepackage{dcolumn}
\usepackage{bm}
\usepackage{verbatim}
\usepackage{multirow}
\usepackage{mathrsfs}
\usepackage{dsfont}
\usepackage{txfonts}
\usepackage{todonotes}

\begin{document}

\title{Majorana bound states in a coupled quantum-dot hybrid-nanowire system}
\author{M.~T.~Deng}
\affiliation{Center for Quantum Devices and Station Q Copenhagen, Niels Bohr Institute, University of Copenhagen, Copenhagen, Denmark}
\affiliation{State Key Laboratory of High Performance Computing, NUDT, Changsha, 410073, China }
\author{S.~Vaitiek\.{e}nas}
\affiliation{Center for Quantum Devices and Station Q Copenhagen, Niels Bohr Institute, University of Copenhagen, Copenhagen, Denmark}
\affiliation{Department of Physics, Freie Universit\"{a}t Berlin, Arnimallee 14, 14195 Berlin, Germany}
\author{E.~B.~Hansen}
\affiliation{Center for Quantum Devices and Station Q Copenhagen, Niels Bohr Institute, University of Copenhagen, Copenhagen, Denmark}
\author{J.~Danon}
\affiliation{Center for Quantum Devices and Station Q Copenhagen, Niels Bohr Institute, University of Copenhagen, Copenhagen, Denmark}
\affiliation{Niels Bohr International Academy, Niels Bohr Institute, University of Copenhagen, Copenhagen, Denmark}
\author{M.~Leijnse}
\affiliation{Center for Quantum Devices and Station Q Copenhagen, Niels Bohr Institute, University of Copenhagen, Copenhagen, Denmark}
\affiliation{Division of Solid State Physics and NanoLund, Lund University, Box. 118, S-22100, Lund, Sweden}
\author{K.~Flensberg}
\affiliation{Center for Quantum Devices and Station Q Copenhagen, Niels Bohr Institute, University of Copenhagen, Copenhagen, Denmark}
\author{J.~Nygård}
\affiliation{Center for Quantum Devices and Station Q Copenhagen, Niels Bohr Institute, University of Copenhagen, Copenhagen, Denmark}
\author{P.~Krogstrup}
\affiliation{Center for Quantum Devices and Station Q Copenhagen, Niels Bohr Institute, University of Copenhagen, Copenhagen, Denmark}
\author{C.~M.~Marcus}
\affiliation{Center for Quantum Devices and Station Q Copenhagen, Niels Bohr Institute, University of Copenhagen, Copenhagen, Denmark}

\date{\today}

\begin{abstract}
\textbf{
Hybrid nanowires combining semiconductor and superconductor materials appear well suited for the creation, detection, and control of Majorana bound states (MBSs). We demonstrate the emergence of MBSs from coalescing Andreev bound states (ABSs) in a hybrid InAs
nanowire with epitaxial Al, using a quantum dot at the end of the nanowire as a spectrometer. Electrostatic gating tuned the nanowire density to a regime of one or a few ABSs. In an applied axial magnetic field, a topological phase emerges in which ABSs move to zero energy and remain there, forming MBSs.We observed hybridization of the MBS with the end-dot bound state, which is in agreement with a numerical model. The ABS/MBS spectra provide parameters that are useful for understanding topological superconductivity in this system.}
\end{abstract}

\maketitle
As condensed-matter analogs of Majorana fermions---particles that are their own antiparticles [1]---Majorana bound states (MBSs)
are anticipated to exhibit non-Abelian exchange statistics, providing a basis for naturally fault-tolerant topological quantum computing
[2-7]. In the past two decades, the list of potential realizations of MBSs has grown from even-denominator fractional quantum Hall states [8] and p-wave superconductors [9] to topological insulator-superconductor hybrid systems [10], semiconductor-superconductor (Sm-S) hybrid nanowire systems [11-21], and artificially engineered Kitaev chains [22-24]. Sm-S hybrid systems have received particular attention because of ease of realization and a high degree of experimental control. Experimental signatures of MBS in Sm-S systems have been reported [25-29], typically consisting of zero-bias conductance peaks in tunneling spectra appearing at finite magnetic field.

\begin{figure}[ht]
\centering \includegraphics[width=8.5cm]{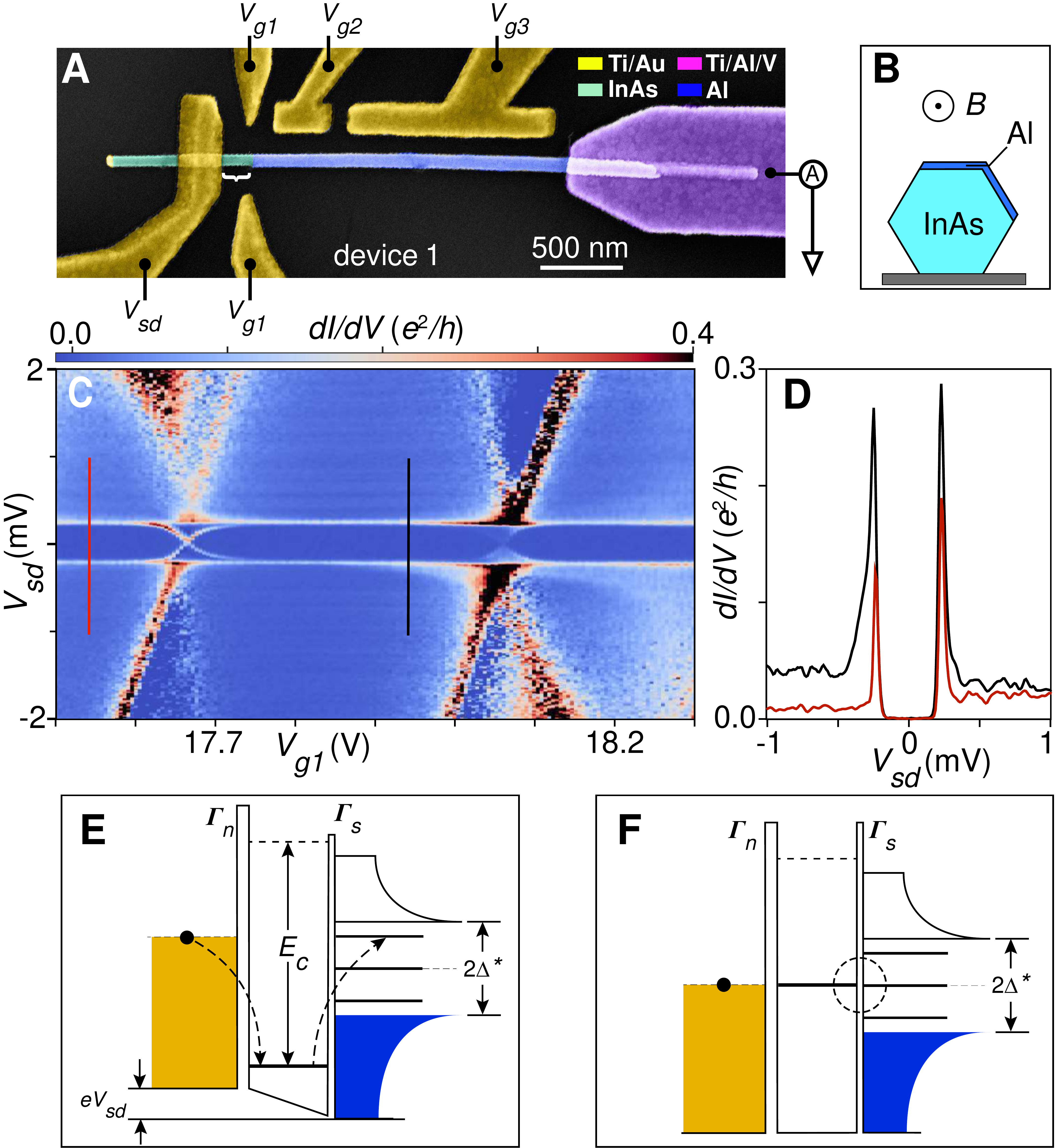}
\caption{\label{Fig1} \textbf{Epitaxial hybrid nanowire with end dot.} \textbf{A,} Scanning electron micrograph (SEM) of device \emph{\textbf{1}}, with false color representing different materials. The white brace indicates the location of a natively formed quantum dot. \textbf{B,} Schematic cross-sectional view of the nanowire. The epitaxial Al shell (blue) was grown on two facets of the hexagonal InAs core (cyan), with a thickness of $\sim$~10 nm. The applied magnetic field is parallel to the nanowire in most cases. \textbf{C,} Differential conductance measured for device \emph{\textbf{1}} as a function of applied source-drain bias voltage, $V_{sd}$, and the voltage $V_{g1}$ on gate $g1$. A Coulomb diamond pattern and a low-conductance gap through the valleys can be seen. \textbf{D,} Line-cuts of the conductance, taken from \textbf{C}, indicated by  red and black lines. \textbf{E, F,} Schematic views of two different dot-wire configurations of the device. \textbf{E} illustrates the elastic cotunneling process in the Coulomb-blockade regime, while \textbf{F} shows how a quantum-dot level can hybridize with the subgap states in the nanowire when it is tuned to resonance.}
\end{figure}

In a confined normal conductor-superconductor system, Andreev reflection will give rise to discrete electron-hole states below the superconducting gap---Andreev bound states (ABSs). Given the connection between superconducting proximity effect and ABSs, zero-energy MBSs in Sm-S hybrid nanowires can be understood as a robust merging of ABSs at zero energy, thanks in part to the presence of strong spin-orbit interaction (SOI) [11-13, 15, 16]. However, not all zero-energy ABSs are MBSs. For instance, in the non-topological or trivial phase, ABSs can move to zero energy at a particular Zeeman field, giving rise to a zero bias conductance peak, and then split again at higher fields, indicating a switch of fermion parity [30]. On the other hand, zero-energy MBSs in short wires may also split as a function of chemical potential or Zeeman field [14]. In this case, the difference between topological MBSs in a finite-length wire and trivial ABSs is whether the states are localized at the wire ends or not [17]. We will use the term ``MBSs'' to refer to ABSs that are to a large degree localized at the wire ends and would evolve into true topological MBSs as the wire becomes longer. We also will use the term ``topological phase in a finite-length wire'' to refer to the regime in which MBS appears. The similarities between trivial ABS zero-energy crossings and MBS in a finite-length wire can be subtle [13, 15, 16, 30, 31]. Several obstacles have prevented a detailed experimental study of the ABS-MBS relation to date, including a soft proximity induced gap [18], the difficulty of tuning the chemical potential of the hybrid nanowire, and disorder in the wire and tunneling barrier.

In this work, we investigate MBSs and their emergence from coalescing ABSs, using tunneling spectroscopy through quantum dots at the end of epitaxial hybrid Sm-S nanowires. We observe gate-controlled hybridization of the MBSs with the bound state in the end dot, finding excellent agreement between experiment and numerical models.  The epitaxial Sm-S interface induces a hard superconducting gap [32, 33], while the partial coverage by the epitaxial superconductor allows tuning of the chemical potential, and yields a high critical field [34], both crucial for realizing MBSs.

\paragraph*{\textbf{Hybrid nanowire with end dot}}
\vspace{8pt}
Our devices were made of epitaxial InAs/Al nanowires (Fig. 1A) [32]. Wurtzite InAs nanowires were first grown to a length of 5 to 10~$\mu$m by means of molecular beam epitaxy, followed by low-temperature epitaxial growth of Al. Two or three facets of the hexagonal InAs core were covered by Al (Fig. 1B) [32]. The nanowires were then deposited onto a degenerately doped silicon/silicon oxide substrate. Transene-D Al etch was used to selectively remove the Al from the end of the wire, which was then contacted by titanium/gold (Ti/Au, 5/100 nm), forming a normal (non-superconducting) metal lead. Five devices were investigated. Data from four devices, denoted \emph{\textbf{1}} to \emph{\textbf{4}}, are reported in the main text, and data from a fifth device, denoted \emph{\textbf{5}}, are reported in Ref. [35] (supplementary materials). For device \emph{\textbf{1}}, the un-etched end of the nanowire section was contacted by titanium/aluminum/vanadium (Ti/Al/V, 5/20/70 nm), and global back gate and local side gates were used to control the electron density in the wire. A quantum dot was formed in the 150-nm bare InAs wire segment between the Ti/Au normal contact and the epitaxial Al shell, owing to disorder or band-bending [33]. Fabrication details for the other devices, each slightly different, are given in Ref. [35]. Micrographs of all devices accompany transport data. Except where noted, the magnetic field B was applied parallel to the nanowire axis by using a three-axis vector magnet. Transport measurements were performed by using standard ac lock-in techniques in a dilution
refrigerator, with a base temperature of 20 mK.

Differential conductance measured for device \emph{\textbf{1}} is shown in Fig. 1C as a function of source-drain voltage,  $V_{sd}$, between the normal and superconducting leads, and the voltage, $V_{g1}$, on gate \emph{g}1. The height (in $V_{sd}$) of the Coulomb-blockade diamond yields an end-dot charging energy $E_c\sim$ 6~meV. Because $E_c$ is larger than the superconductor gap, single-electron cotunneling dominates transport in Coulomb-blockade valleys. In this regime, the dot acts effectively as a single barrier and can be used as a tunneling spectrometer for the wire (Fig. 1E). On the other hand, when the dot is tuned onto a Coulomb peak (Fig. 1F), hybridization occurs between the dot and wire states [36]. We first discuss cotunneling spectra away from resonance then investigate dot-wire interaction when the dot is on resonance with ABSs and MBSs in the wire.

\paragraph*{\textbf{Weak dot-wire coupling}}

\begin{figure*}[ht]
\centering \includegraphics[width=16cm]{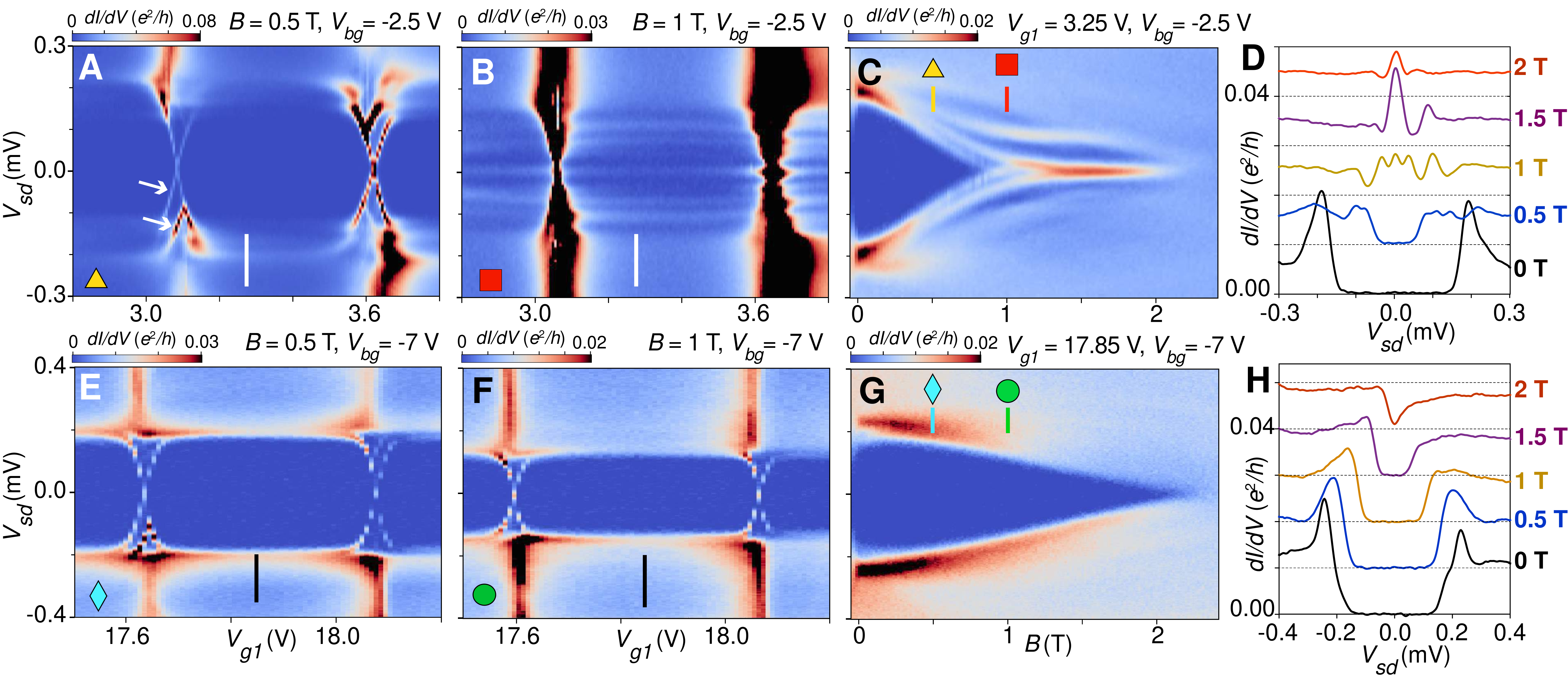}
\caption{\label{Fig2} \textbf{Tunneling spectra for large and zero ABS density.} \textbf{A,} Differential conductance measured for device \emph{\textbf{1}} as a function of $V_{sd}$ and $V_{g1}$, measured at $B=0.5$~T and $V_{bg} = -2.5$~V, $V_{g2, g3}=-10$~V. The white arrows indicate Zeeman split dot levels. \textbf{B,} The same as \textbf{A}, but at $B=1$~T. ABSs can be clearly identified below the superconducting gap. \textbf{C,} Differential conductance as a function of $V_{sd}$ and $B$ ($B$-$V_{sd}$ sweep), measured at the gate voltage indicated by the white lines in \textbf{A} and \textbf{B}. The triangle and square indicate at which field \textbf{A} and \textbf{B} are measured.  Blurring of data in narrow $B<0$ range is due to heating caused by sweeping field away from zero. \textbf{D,} Line-cut taken from \textbf{C} at various $B$ values. Lines are offset by 0.01~$e^2/h$ each for clarity. The conductance peaks below the superconducting gap indicate that the wire is in a subgap-state-rich regime. A well defined zero-bias peak can be seen at high field. \textbf{E-H,} Similar to \textbf{A-D}, but measured at $V_{bg}=-7$~V and $V_{g2, g3}=-10$~V, and with \textbf{G} measured at the gate voltage indicated by the black lines in \textbf{E} and \textbf{F}. The diamond and circle indicate at which field \textbf{E} and \textbf{F} are measured. Here, a hard superconducting gap is clearly seen, with a critical magnetic field $B_c$ up to $\sim$~2.2~T. No subgap structure is observed across the full range of field, 0 -- 2 T.}
\end{figure*}

\vspace{8pt}
A hard proximity-induced superconducting gap, marked by vanishing conductance below coherence peaks, can be seen in cotunneling transport through Coulomb blockade valleys of the end dot (Fig. 1D). The width of the gap in bias voltage is given by $2\Delta^*/e$, where $\Delta^*$ is the effective superconducting gap, defined phenomenologically by the bias voltage at which the quasiparticle continuum appears. The value of $\Delta^*$ for device \emph{\textbf{1}} is found to be 220~$\mu$eV (for devices \emph{\textbf{2}}, \emph{\textbf{3}}, and \emph{\textbf{4}}, $\Delta^*\sim$ 250 to 270~$\mu$eV), which is somewhat larger than measured previously in either epitaxial [33] or evaporated hybrid devices [27, 37]. The measured gap is consistent with values for evaporated ultra-thin Al films in the literature [38].

Tunneling conductance ($dI/dV_{sd}$) for device \emph{\textbf{1}}, as a function of $V_{g1}$ and $V_{sd}$, spanning three Coulomb blockade valleys is shown in Fig. 2 for two values of back-gate voltage $V_{bg}$, which is applied uniformly to the device by using a conductive Si substrate separated by a 200-nm oxide layer. To compensate the effect of $V_{bg}$ on the conductance of the end dot, the voltage $V_{g1}$, on the gate near the end dot is simultaneously swept by a small amount during the back-gate sweep. Other gates are grounded. At less negative back-gate voltage ($V_{bg} = -2.5$ V), several subgap conductance peaks are seen at $B=1$~T, including one at zero bias. We attribute these peaks, which run through consecutive Coulomb valleys, to ABSs in the finite-length wire. The magnetic field dependence of the spectrum is shown in Fig. 2, C and D: subgap states lie close to the superconducting gap at zero field and move to lower energies as B increases. Some of the lower-energy subgap states merge at zero energy, forming a narrow zero-bias peak spanning the range from 1 to 2~T. At more negative back-gate voltage, $V_{bg} = -7$  V, dot-independent subgap structure is absent (Fig. 2, E to H); only a hard superconducting gap is seen throughout the field range of 0 to 2~T. The back-gate dependence on the number of ABSs in the gap demonstrates that the chemical potential of the wire can be controlled with the superconductor shell present.

The zero-field effective gap $\Delta^*$ in the regime with high ABS density is $\sim$200~$\mu$eV, which is distinctly smaller than the 220~$\mu$eV gap seen in the no-ABS regime. This is because the phenomenological $\Delta^*$ in the high-ABS density regime is mainly determined by the energy of the cluster of ABSs, yielding what is usually referred to as the induced gap $\Delta_{ind}$. When there are no states in the wire, $\Delta^*$ is set by the gap of the Al shell, denoted $\Delta$.

\begin{figure*}[ht]
\centering \includegraphics[width=16cm]{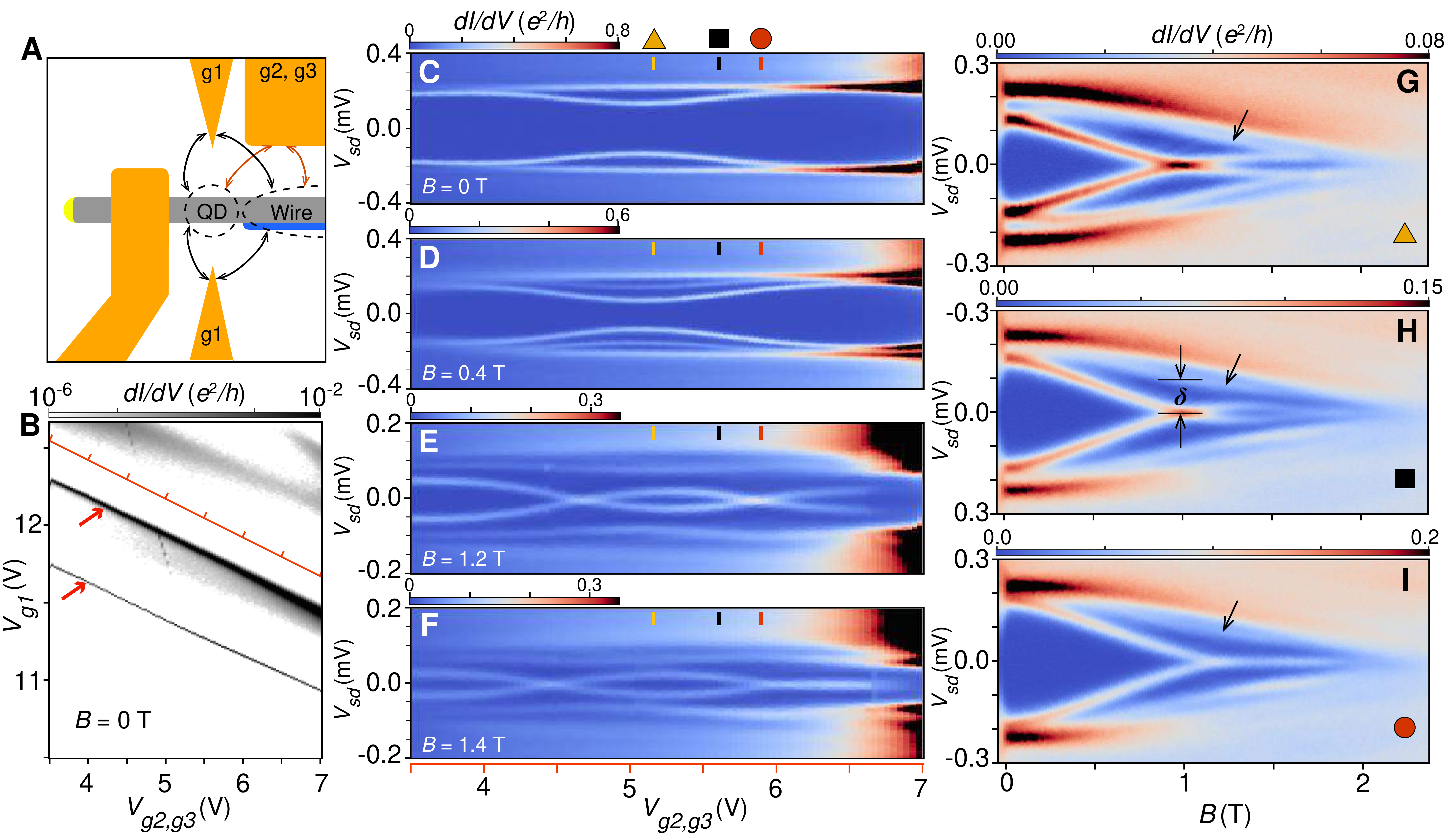}
\caption{\label{Fig3} \textbf {Tunneling spectra for intermediate density in few-ABS regime.} \textbf{A,} A schematic of device \emph{\textbf{1}} showing the gating configuration for a combined gate voltage sweep. $g1$ and $g2, g3$ are capacitively coupled to both the dot and the nanowire. \textbf{B,} Conductance measured at $V_{sd}=0$~mV, $V_{bg}=-7$~V and $B=0$, as a function of $V_{g1}$ and $V_{g2, g3}$ (the gate map). Note that $g2$ and $g3$ are connected to the same voltage source. The high-conductance lines indicated by red arrows are the resonant levels in the end-dot. The dot can be used as a cotunneling spectrometer if the gate sweeping is kept inside the Coulomb blockade valley and parallel to the resonant level. \textbf{C-F,} Tunneling spectra at various magnetic fields as a function of the combined gate voltage, measured along the red line in \textbf{B}. The energy of the ABSs is strongly dependent on gate voltages. \textbf{G-I,} $B$-$V_{sd}$ sweeps at different gate voltages, corresponding to the triangle, square and circle in \textbf{C-F}, respectively. Depending on gate voltages, the ABSs in the wire show different magnetic field evolution, from a splitting behaviour (\textbf{G}) to non-splitting behaviour (\textbf{I}). Arrows in \textbf{G-I} indicate the first excited ABSs, and $\delta$ in \textbf{H} is defined as the residual gap, i.e., the energy of the first excited state around topological phase transition, due to the finite-size effect.}
\end{figure*}

Between the regimes of high ABS density and zero ABS density, one can find, by adjusting back and local gates, a low-density ABS regime in which only one or a few subgap modes are present. In this intermediate density regime, ABSs can be readily probed with tunneling spectroscopy, without softening the gap with numerous quasicontinuous subgap states. To prevent end-dot states from mixing with ABSs in the wire, two gate voltages, one at the junction and one along the wire, were swept together so as to compensate for capacitive cross-coupling (Fig. 3A). In this way, either the end-dot chemical potential $\mu_{\rm dot}$ or the wire chemical potential $\mu_{\rm wire}$ could be swept, with the other held fixed. A two-dimensional plot of zero-bias conductance as a function of $V_{g1}$ and $V_{g2,g3}$ (fixing $V_{g2} = V_{g3}$) in Fig. 3B shows isopotential lines for the end dot as a diagonal Coulomb blockade peak-ridge (red arrows). The slope of this ridge determines how to compensate the wire gates ($V_{g2,g3}$) with the junction gate ($V_{g1}$). Data can then be taken in the cotunneling regime for an effectively constant $\mu_{\rm dot}$. Tunneling spectra measured along the red line in Fig. 3B at various fields are shown in Fig.3, C to F. A pair of ABSs that moves with $\mu_{\rm wire}$ can be seen at $B = 0$ (Fig. 3C). The spectrum is symmetric around zero $V_{sd}$, reflecting particle-hole symmetry. The minimum energy of the ABS is $\zeta=130~\mu$eV, which is smaller than the effective gap $\Delta^*=220~\mu$eV. The pair of ABSs splits into two pairs when the applied magnetic field lifts the spin degeneracy, visible above $B = 0.4$~T (Fig. 3D). The low field splitting corresponds to an effective \emph{g}-factor, \emph{g*}$\sim$4 (the \emph{g*}-factor estimated from the ABS-energy/magnetic field slope may differ considerably from the intrinsic \emph{g*}-factor). At higher magnetic fields, the inward ABSs cross at zero and reopen, forming a characteristic oscillatory pattern (Fig. 3, E and F). The gap reopening at more positive $V_{g2,g3}$ is relatively slow, leading to a single zero-bias peak in the range of $V_{g2,g3}\sim 5.8$--$7$~V (Fig. 3F).

The magnetic field dependence of the ABS spectrum near the ABS energy minimum is shown in Fig. 3G. The evolution of the ABSs can be clearly followed: They split at low field, the inner ABSs merge around $B = 1$~T, they split again at higher fields, and the resplit ABSs merge with the higher-energy ABSs above $B = 1.7$~T. Here, the emergence of a zero-bias peak and its splitting is qualitatively similar to the observations reported in Refs. [27, 30]. However, the
$B$-dependent ABS spectrum at more positive gate voltage (Fig. 3H) shows a merging-splitting-merging behavior, giving rise to an eye-shaped loop between 1 and 2~T. At even more positive gate voltage (Fig. 3I), the spectrum displays an unsplit zero-bias peak from 1.1 to 2~T. The first excited ABSs in Fig. 3, G to I, are still visible at a high magnetic field---for instance, as marked at B = 1.2 T in Fig. 3, H and I. Qualitatively, the lowest-energy ABSs in Fig. 3, H and I, tend to split after crossing but are pushed back by the first excited ABSs, resulting in either a narrow splitting or an unsplit zero-bias peak. The measurements in Fig. 3, C to I, were taken in an even Coulomb valley of the end dot, but the qualitative behavior does not depend on end-dot parity. Similar results measured in an odd valley of the end dot are provided in Ref. [35].

\begin{figure*}[ht]
\centering \includegraphics[width=16cm]{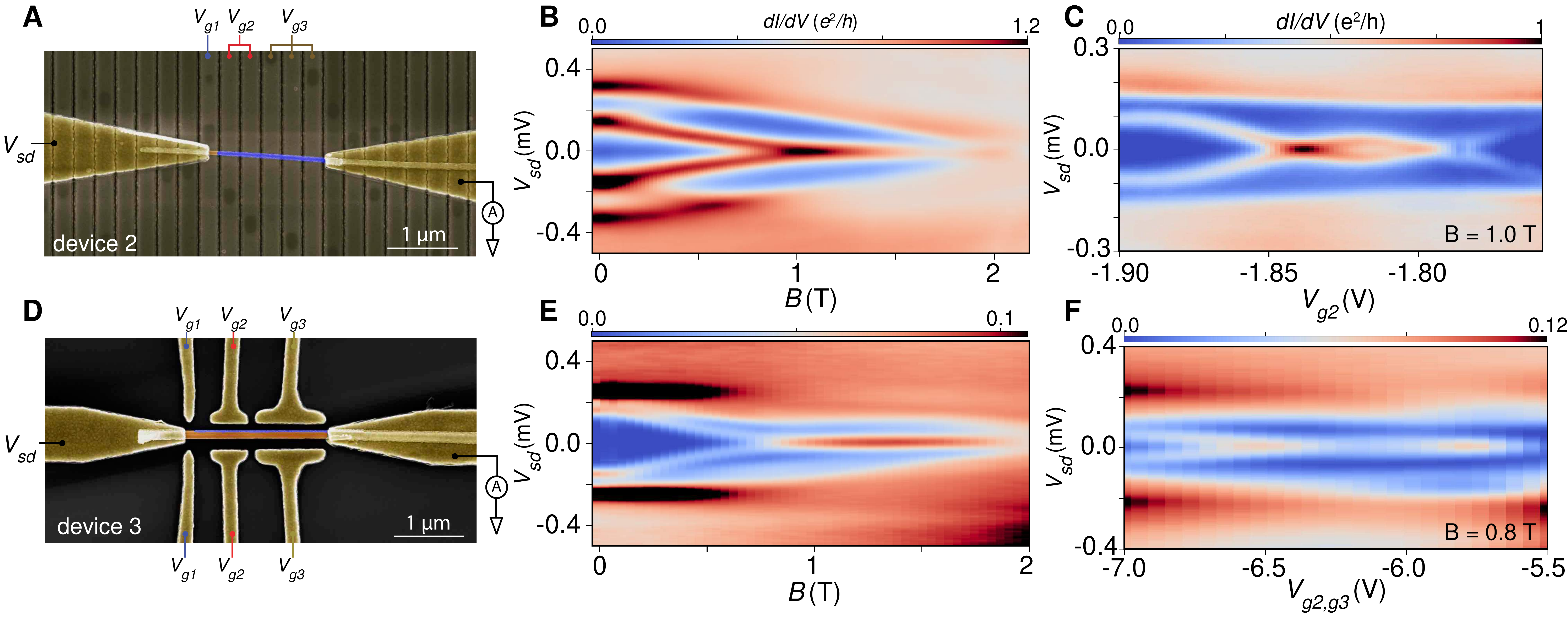}
\caption{\label{Fig4} \textbf{Stable zero energy states measured on other devices}. \textbf{A} SEM images of device \emph{\textbf{2}}, in which local bottom gates are used. The hybrid wire section is 1.5~$\mu$m long. \textbf{D} SEM image of device \emph{\textbf{3}}, with the hybrid wire section length around 2~$\mu$m long. \textbf{B,E} Subgap states evolution in magnetic field, measured on device \emph{\textbf{2}} and device \emph{\textbf{3}}, respectively. In both plots, stable zero energy states arising from a pair of ABSs can be seen. \textbf{B} is measured at $V_{g1}=-600$~mV, $V_{g2}=-1840$~mV, $V_{g3}=5$~V, and \textbf{E} is measured at $V_{g1}=3720$~mV, $V_{g2}=V_{g3}=-5850$~mV, $V_{bg}=-8$~V. \textbf{C,F} Gate voltage dependence measurements of subgap states for device \emph{\textbf{2}} and device \emph{\textbf{3}}, respectively. Both measurements are taken by following the isopotential lines of the hybrid wires in one of their end-dot Coulomb blockade valleys.}
\end{figure*}

The different field dependences of the subgap states---either crossing zero or sticking at zero---can be understood as reflecting a transition from ABS to MBS [14, 17]. For the ABSs in the regime of $V_{g2,g3}<5.8$~V, their crossing in Zeeman field is a signature of parity switching, similar to ABSs
in a quantum dot, such as investigated in Ref. [30]. In contrast, behavior in the range of $V_{g2,g3}\sim 5.8 $--$ 7$~V indicates that the system is in the topologically nontrivial regime, with MBS levels that stick to zero as the magnetic field increases. In a finite-size wire, SOI induces anticrossings between
discrete ABSs, thus pushing levels to zero, preventing further splitting. We ascribe the differences in the qualitative behavior in Fig. 3, G to I, to state-dependent SOI-induced anticrossings, which depend on gate voltage. The excited ABS in Fig. 3G and the ones in Fig. 3, H and I, are presumably not the same state, but belong to different subgap modes (investigated in detail in Ref. [35]).

For a long wire, the topological phase transition is marked by a complete closing and reopening of a gap to the continuum, with a single discrete state remaining at zero energy after the reopening. For a finite wire, the continuum is replaced by a set of discrete ABSs, and at the transition where a single state becomes pinned near zero energy, there remains a finite gap  $\delta$ to the first discrete excited state. At this transition point (where the gap of the corresponding infinite system would close and its spectrum would be linear), $E_k=R\alpha |k|$, where $R$ is a renormalization factor due to the strong coupling between the semiconducting wire and its superconducting shell [35], $\alpha$ is the spin-orbit coupling strength, and $k$ is the electron wave vector. From this relation, we can connect $\delta$ to the ratio $L/\xi$ as $L/\xi\approx R\pi\Delta^\prime/\delta$, where $L$ is the separation between Majoranas (the wire length in the clean limit), $\xi$ is the effective superconducting coherence length near the topological phase transition, and $\Delta^\prime$ is the effective gap near the phase transition point (the derivation and more details are available in Refs. [35, 39, 43]). The ratio $L/\xi$ is the dimensionless length of the topological wire segment. We estimate from Fig. 3H $\delta \sim100~\mu$eV, $\Delta^\prime\sim180~\mu$eV and $R\sim0.4$, yielding $L/\xi\sim$ 2.3. We then take values at the field where the ABS reaches zero energy. We can independently estimate $L/\xi$ from the relation $\delta E\approx\delta E_0e^{-L/\xi}$ [14, 17], where $\delta E$ is MBS oscillation amplitude and $\delta E_{0}$ is a pre-factor. If we take the value $\delta E_0 \sim 150~\mu eV$ based on Coulomb peak motion in Ref. [40], we obtain a value $L/\xi\sim$ 1.3 by using the subgap state splitting energy $\delta E \sim 40~\mu$eV at $B\sim1.3$ T from Fig. 3H. We speculate that the discrepancy in estimates of $L/\xi$ may be attributed to a smaller value of $\delta E_{0}$ in Ref. [40] as compared with the $\delta E_{0}$ in this device, perhaps arising from differences in gate-tuned electron density compared with the Coulomb blockade devices in Ref. [40].

Subgap-state evolution in applied magnetic field and gate voltage for devices \emph{\textbf{2}} and \emph{\textbf{3}} are shown in Fig. 4. For device \emph{\textbf{2}}, which has a device length of $\sim 1.5$~$\mu$m, subgap state reaches zero energy at $\sim$ 0.9 T and persists to 2 T. Using the finite-size gap near the phase transition point in Fig. 4B, we can extract $L/\xi\sim 1.7$ ($\delta\sim\Delta^\prime\sim170~\mu$eV, $R\sim$~0.56). For device \emph{\textbf{3}}, with a $\sim$2~$\mu$m hybrid nanowire, rigid zero energy states are shown in both magnetic field and gate voltage-dependence measurements. Here, a value of $L/\xi\sim 3.4$ ($\Delta^\prime\sim 230~\mu$eV, $\delta\sim100~\mu$eV, $R\sim$~0.47) can be estimated from Fig. 4E.

The obtained values for $L/\xi$ evidently do not reflect the lithographic device length. For instance, taking $\xi\sim260$~nm from Ref. [40] (where $\xi$ is fit from the wire-length dependence by using similar nanowires), yields lengths of 0.6, 0.45, and 0.9~$\mu$m for devices \emph{\textbf{1}}, \emph{\textbf{2}}, and \emph{\textbf{3}}, respectively, in each case shorter than the lithographic length. This discrepancy is presumably the result of disorder or material defects that create a topological region shorter than the full wire.

\paragraph*{\textbf{Resonant dot-MBS coupling}}

\vspace{8pt}

\begin{figure}[h]
\centering \includegraphics[width=8.5cm]{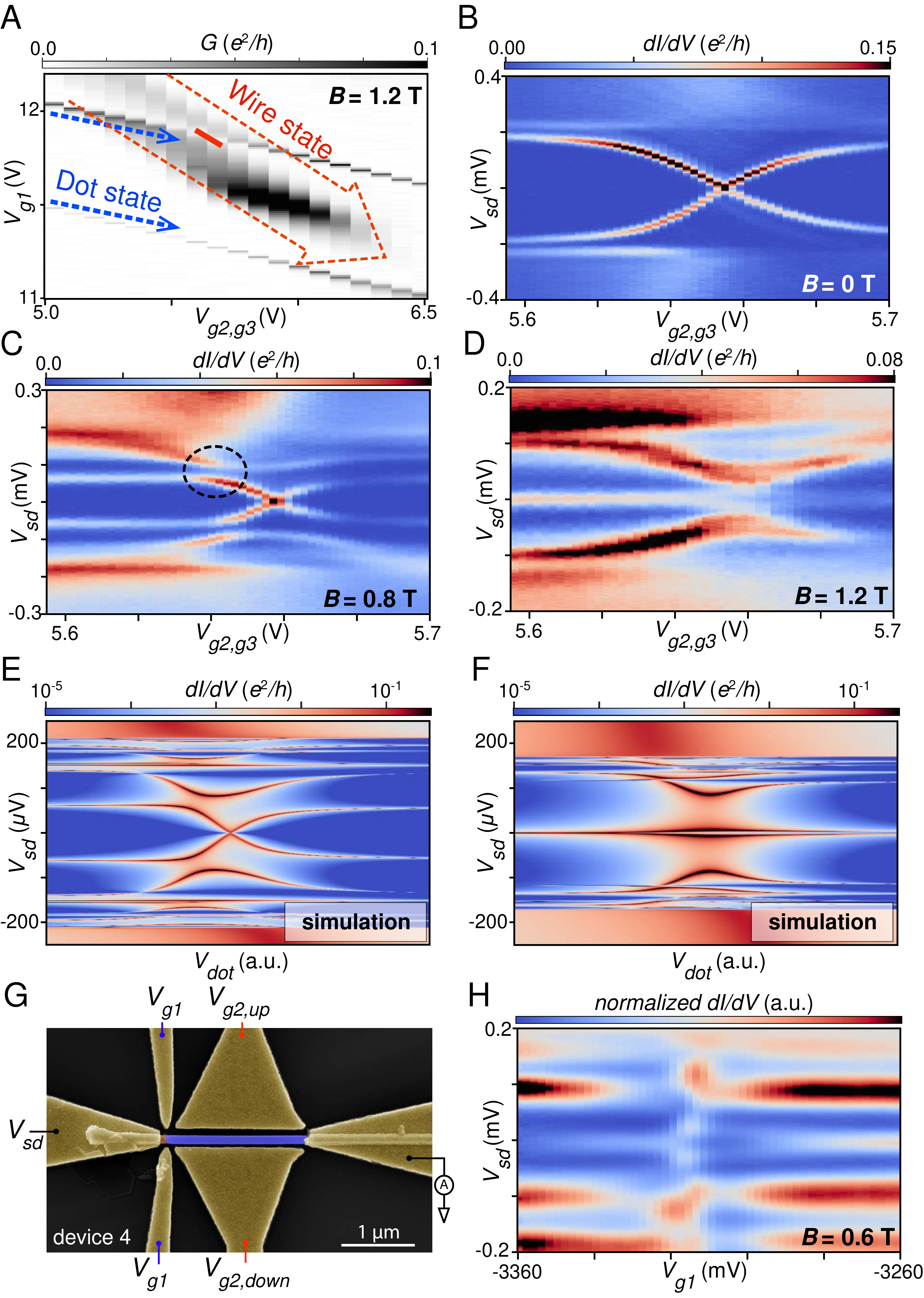}
\caption{\label{Fig5} \textbf{Resonant coupling of wire subgap states and dot states}. \textbf{A,} A gate map similar to Fig.~\ref{Fig3}b, but taken at $B=1.2$~T. The blue dashed lines denote the dot isopotential and red dashed arrow the wire isopotential sweeping directions. \textbf{B-D,} Differential conductance measured across one of the resonant dot levels, along the solid red line in \textbf{A}, at $B=0$, 0.8~T, and 1.2~T. The dashed circle in \textbf{C} indicates an anticrossing between the dot state and a wire state. In \textbf{D} there is a pronounced zero-bias peak-splitting at the dot resonance. \textbf{E, F,} Simulated differential conductance through a dot-hybrid wire system, as a function of bias voltage and dot chemical potential $\mu_{\rm dot}$, for different Zeeman splittings in the wire. The zero-bias peak splitting at the dot resonance also appears in \textbf{F}. \textbf{G,} SEM image of device \emph{\textbf{4}}. \textbf{H,} Normalized conductance in the MBS-dot hybridization region of device \emph{\textbf{4}}. Again the zero energy MBS state is split when it crosses the end dot resonant level.}
\end{figure}

We next examined the interaction of wire states with bound states in the end dot. For all data above, the wire states were probed via cotunneling through the end dot (Fig. 1E), actively keeping the end dot in the middle of a Coulomb valley. By separately controlling $\mu_{\rm dot}$ and $\mu_{\rm wire}$, we can also tune the local gates so that a wire state is at resonance with a Coulomb peak of the dot (Fig. 1F). As seen in the gate map for device \emph{\textbf{1}} shown in Fig. 5A, besides the dot resonant level, there is a MBS-induced blurred trace (Fig. 5A, red dashed arrow). Tunneling spectra at various magnetic fields where the MBS and end-dot state align are shown in Fig. 5, B to D (traces with a larger range are provided in Ref. [35]). The gate sweep is now along a wire isopotential (Fig. 5A, red solid line) as opposed to an end-dot isopotential as in Fig. 3. The dot state crosses zero energy around $V_{g2,g3} = 5.66$~V, at which the dot switches its fermion parity. At $B = 0$ (Fig. 5B) and at $B = 0.8$~T (Fig. 5C), we clearly see this crossing in the spectrum. There is a pronounced anticrossing between the dot state and the wire state in Fig. 5C (indicated by the dashed circle). Figure 5D looks qualitatively different: A zero-bias peak is visible when the dot is off resonance (which is extended from the MBS in Fig. 3H), and this peak splits when
the dot level comes close to zero. In this case, no zero-crossing is observed.

The dot-wire interaction observed in Fig. 5D can be understood in terms of leakage of the MBS into the dot when the dot is on resonance [41]. The energy splitting of a pair of MBSs is given by $\delta E \propto|\sin { \left({ k }_{ F }L\right) }{ e }^{ -L/\xi }|$  (where $k_F$ is the effective Fermi wave vector). In Fig. 5D, this splitting is initially small, when the dot is off resonance and coupling of the MBSs to the dot states is suppressed by Coulomb blockade. For a finite size wire, this implies that $\sin { \left({ k }_{\rm F }L\right) }\sim 0$ at that particular tuning. As the dot level comes closer to the resonant point, the nearby MBS partially leaks into the dot, which changes the details of the MBSs wave function (the numerical study on the wave-function distribution is provided in Ref. [35]). This can change the effective $k_{\rm F}L$ in $\delta E$, which causes the zero-bias peak to split at resonance. Numerical simulations of the conductance spectrum of the coupled dot-MBS (Fig. 5, E and F) show good qualitative agreement with the experimental data, both in the trivial superconducting regime (Fig. 5, C and E) and in the topological superconducting phase (Fig. 5, D and F). Similar zero-bias peak splitting in another coupled dot-MBS device (device \emph{\textbf{4}}) is shown in Fig. 5H. To enhance image visibility, conductance values in Fig. 5H are normalized by the conductance at  $V_{sd}=0.2$~mV at the corresponding gate voltage.

\begin{figure}[h]
\centering \includegraphics[width=8.5cm]{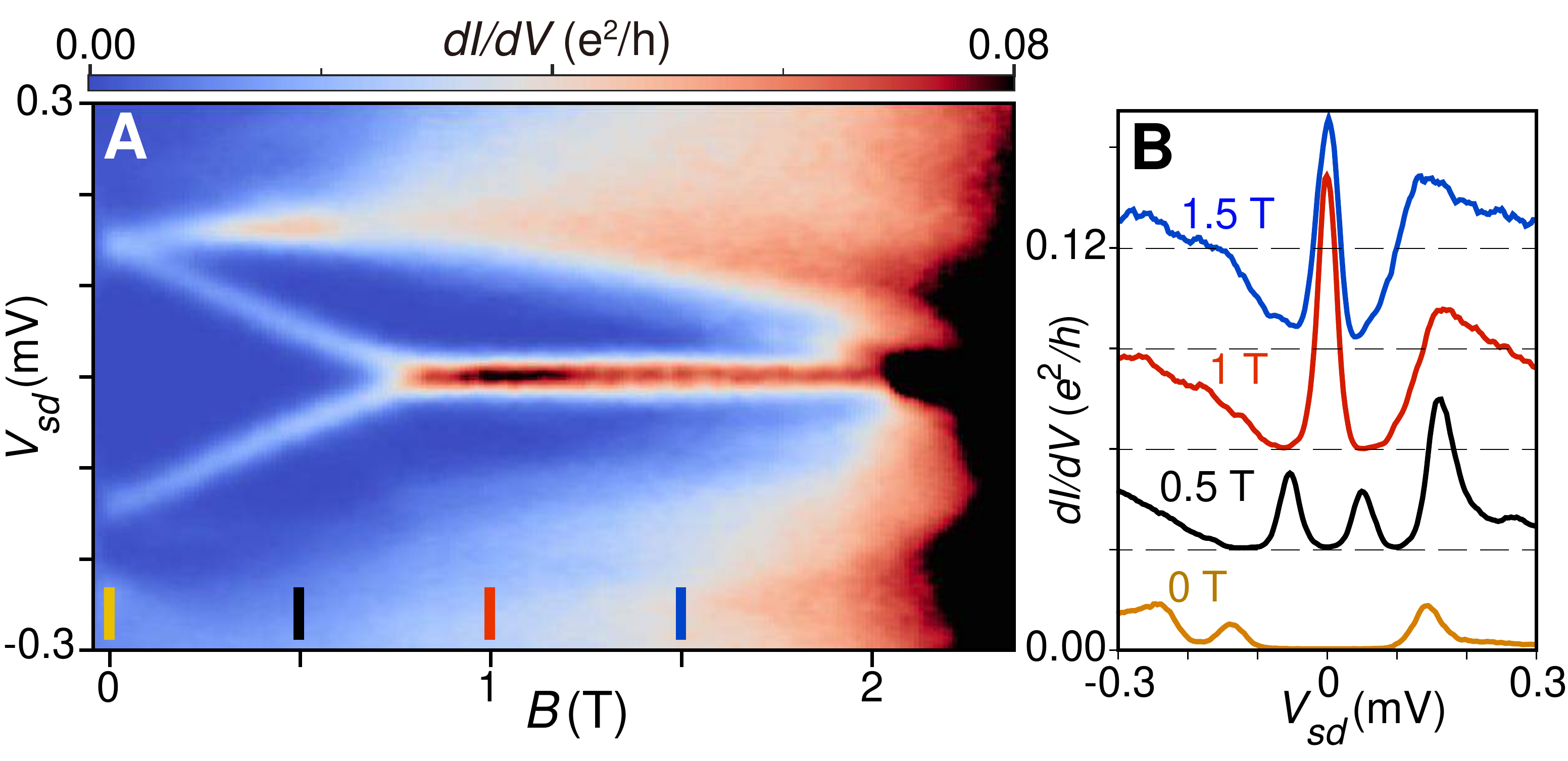}
\caption{\label{Fig6} \textbf{Tunneling spectrum for resonant dot-wire coupling.} \textbf{A,} $B$-$V_{sd}$ sweep at $V_{bg}=-8.5$~V, $V_{g1}=22$~V, and $V_{g2}=V_{g3}= -10$~V. \textbf{B,} Differential conductance line-cut plots taken from \textbf{A} at various $B$ values. At this gate configuration, a pronounced zero-bias conductance peak emerges around $B=0.75$~T and persists above $B=2$~T without splitting. The intensity of the zero-bias peak is relative higher than other finite-energy ABSs, and even higher than the Al superconducting coherence peaks. The background conductance is almost zero even at $B=1$~T, indicating the induced gap is still a hard gap after the phase transition.}
\end{figure}

Last, we examined the magnetic field evolution of the subgap states in the strong dot-wire coupling regime, in which dot and wire states cannot be separated. Shown in Fig. 6 is the evolution with field of the spectral features of the dot-wire system measured for device \emph{\textbf{1}}, with two ABSs merging at $B = 0.75$~T into a stable zero-bias peak that remains up to $B = 2$~T. The effective \emph{g*}-factor that can be deduced from the inward ABS branches is $\sim$6. The conductance at the base of the zero-bias peak is almost zero even at $B = 1$~T, indicating a hard superconducting gap also after the topological phase transition. Related measurements are shown in Ref. [35].

The long field range and intensity of the zero-bias peak in Fig. 6 can be understood as arising from the hybridization of the MBS with the end-dot state. In the strong coupling regime, MBS can partially reside at the end dot, making the effective length of the wire longer than in Fig. 3I. The MBS wave function has larger amplitude at the wire end, where the dot couples, than either finite-energy ABSs or states in the Al shell. This leads to a relatively higher conductance peak at zero energy and makes the excited states and the Al shell superconducting coherence peaks almost invisible [13]. The long field range of the zero-bias peak in Fig. 6 (also Fig. 3I) may also be enhanced by electrostatic effects that depend on magnetic field [14, 19].

Our measurements have revealed how the ABSs in a hybrid superconductor-semiconductor nanowire evolve into MBSs as a function of field and gate voltage.



\vspace{12pt}
\textbf{Acknowledgements}
We thank Ramon Aguado, Sven Albrecht, Jason Alicea, Leonid Glazman, Andrew Higginbotham, Bernard van Heck, Thomas Sand Jespersen, Ferdinand Kuemmeth, Roman Lutchyn, and Jens Paaske for valuable discussions, and Veronica Kirsebom, Samuel Moore, Magnus Ravn, Daniel Sherman, Claus Sorensen, Giulio Ungaretti, and Shivendra Upadhyay, for contributions to growth, fabrication, and analysis.~This research was supported by Microsoft Research, Project Q, the Danish National Research Foundation, the Villum Foundation, and the European Commission.~M.~Leijnse acknowledges the Crafoord Foundation and the Swedish Research Council (VR).

\clearpage
\onecolumngrid
\setcounter{figure}{0}
\setcounter{equation}{0}
\setcounter{page}{1}

\newcommand{\bbbone}{\mathchoice {\rm 1\mskip-4mu l} {\rm 1\mskip-4mu l} {\rm
1\mskip-4.5mu l} {\rm 1\mskip-5mu l}}
\newcommand{\Al}{\mathrm{Al}}
\newcommand{\eff}{\mathrm{eff}}
\newcommand*{\p}{\partial}
\renewcommand*{\d}{\;\text{d}}
\renewcommand*{\u}[1]{\underline{#1}}
\newcommand*{\e}{\text{e}}
\renewcommand*{\le}{\left}
\newcommand*{\ri}{\right}
\newcommand*{\twovector}[2]{\left(\begin{array}{c}
    #1\\
    #2
  \end{array}\right)}
\newcommand*{\twomatrix}[4]{\left[\begin{array}{cc}
    #1 & #2\\
    #3 & #4
  \end{array}\right]}
\newcommand{\sigmax}{\left(\begin{array}{cc}
    0 & 1\\
    1 & 0
  \end{array}\right)}
\newcommand{\sigmay}{\left(\begin{array}{cc}
    0 & -i\\
    i & 0
  \end{array}\right)}

\section{Supplementary Material:}
\section{Majorana bound states in a coupled quantum-dot hybrid-nanowire system}

\vspace{0.1in}

This supplementary material includes:\\

\textbf{Supplementary Texts}
\begin{enumerate}
  \item Devices fabrication details,
  \item Measurements in high density many-ABS regime,
  \item MBS dot-occupancy dependent measurements,
  \item Longitudinal ABS modes from the same subband,
  \item Zero-bias peaks in strong dot-wire coupling regime,
  \item Magnetic field orientation measurements of subgap states,
  \item Simulation of Majorana-dot interaction,
  \item Extracting parameters from the observed spectrum.
\end{enumerate}

\vspace{10pt}
\textbf{Figures S1 to S11.}

\newpage
\vspace{20pt}
\paragraph*{\textbf{Devices fabrication details}}

All the devices in this article were made of epitaxial InAs-Al nanowires grown by molecular beam epitaxy. The InAs nanowires in devices \emph{\textbf{1-5}} were grown along $\textless111\textgreater$ direction with six facets, and two/three of them were covered by low-temperature epitaxial grown Al. The Al film is on two facets of the InAs core with a thickness $\sim10$~nm for the nanowires in devices \emph{\textbf{1}} and \emph{\textbf{2}}, while the Al shell covers three facets for the nanowires in devices \emph{\textbf{3}} and \emph{\textbf{4}} with a thickness $\sim7$~nm. The nanowire in device \emph{\textbf{5}} was grown along $\textless11\overline{2}\textgreater$ direction with four facets, and one of them was covered by $\sim20$~nm Al. The thickness of Al-shell will influence the measured effective gap $\Delta^*$ at zero magnetic field and the critical field $B_c$.

The nanowires in devices \emph{\textbf{1}} and \emph{\textbf{5}} were transferred to Si/SiO$_{x}$ substrates by a small piece of cleanroom wipe, while the nanowires in devices \emph{\textbf{2-4}} were transferred by a manipulator station using a tungsten needle.

Ti/Al/V leads were used in devices \emph{\textbf{1}} and \emph{\textbf{5}} to contact the hybrid wire segments, and Ti/Au leads were used to contact the etched nanowire segment. DC ion-milling was used to remove the oxide layer before metal deposition. For devices \emph{\textbf{2-4}}, all the leads were made of Ti/Au, and RF ion-milling was used to remove the oxide layer. From the measurement results on these devices, hard proximity effect induced gaps are found in all Ti/Al/V and Ti/Au contacted hybrid nanowire segments, no gap-softening evidence from Ti/Al/V or Ti/Au leads is observed.

In devices \emph{\textbf{1,3,4}},and \emph{\textbf{5}}, local side gates were used to tune chemical potential of the hybrid wire or the end dot. While, local bottom gates were made for device \emph{\textbf{2}}, with $\sim15$~nm thick hafnium dioxide (HfO$_2$) as the dielectric layer which was grown by atomic-layer-deposition.

Except where noted, the measurement results shown in this Supplementary Material are taken on device \emph{\textbf{1}}.

\newpage
\paragraph*{\textbf{Measurements in high density many-ABS regime}}

\begin{figure}
\renewcommand{\thefigure}{S\arabic{figure}}
\centering \includegraphics[width=12cm]{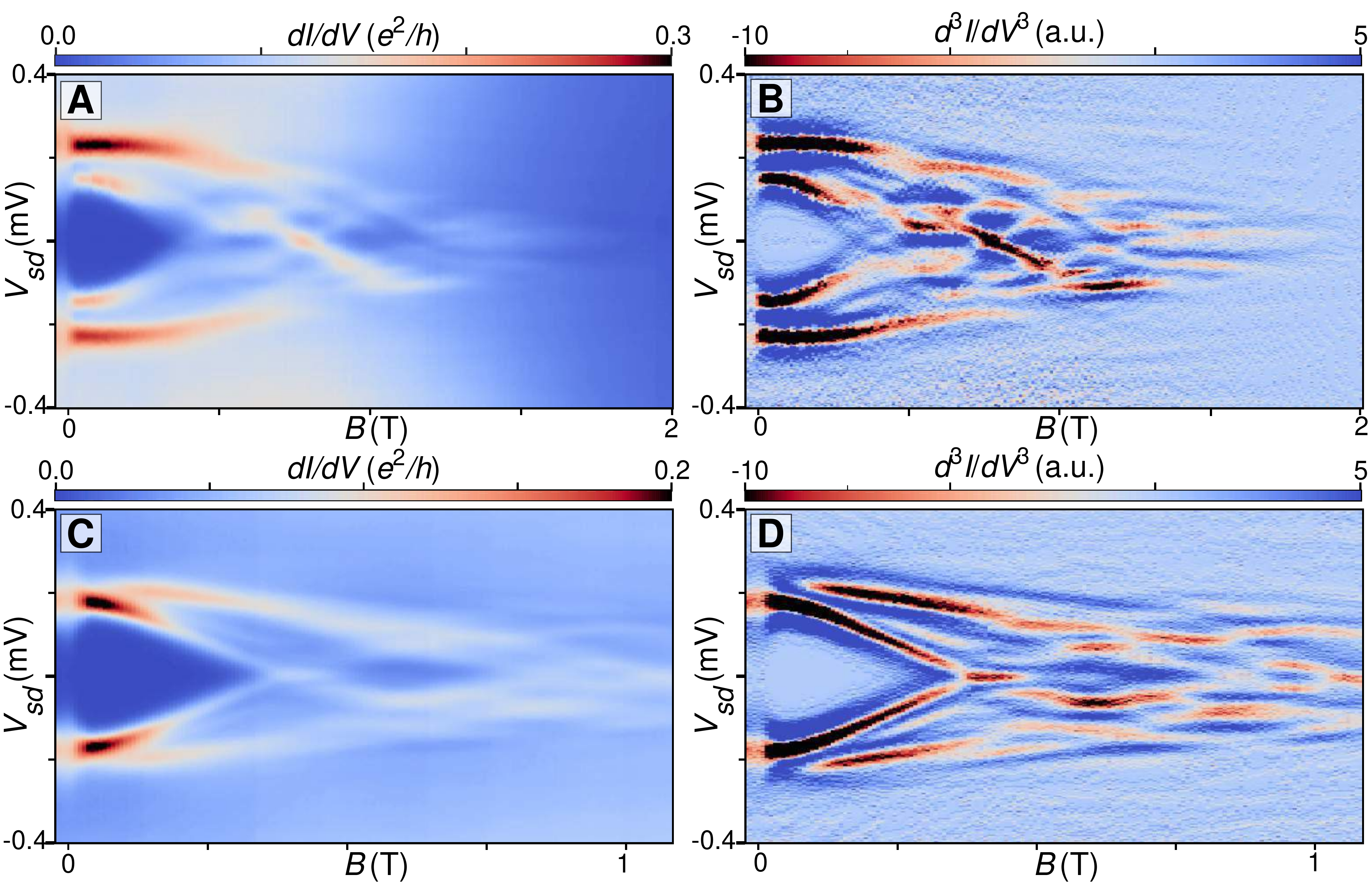}
\caption{\label{FigS1} \textbf{Tunneling spectra in the high density many-ABS regime as a function of magnetic field.} \textbf{a} and \textbf{c}, Differential conductance as a function of source-drain voltage $V_{sd}$ and axial magnetic field $B$, at $V_{bg}=3.9$~V, $V_{g1}=-20$~V for \textbf{a}, and $V_{bg}=4.8$~V, $V_{g1}=-25$~V for \textbf{c}. \textbf{b} and \textbf{d}, The curvature (second derivative) of the conductance, $d^3I/dV^3$, for \textbf{a} and \textbf{c}, respectively. The curvature of conductance helps identifying local maxima of conductance, thus emphasizing spectral features. In this regime of positive $V_{bg}$, many ABSs are visible. Note that ABSs occasionally cross zero energy, creating zero-energy conductance peaks over short ranges of magnetic field.}
\end{figure}

In Fig.~\ref{FigS1} we investigate further the magnetic-field dependence of the subgap states in the high-density many-ABS regime.
%
When $V_{bg}$ is positively biased, the hybrid wire is populated by many ABS modes. To maintain low conductance in the end dot, $V_{g1}$ is set to large negative values. Even in the high density regime the induced superconducting gap is still very hard at zero magnetic field. As the Zeeman field increases, the ABSs branch, merge, and cross zero energy. The superconducting gap is softened at $B\gtrsim 0.5$~T, where it gets populated with discrete, broadened ABSs.
%
It should be emphasized that even in the high-density regime, zero-energy subgap states can emerge and persist over an interval of magnetic field up to 200 mT (see the second derivative of the conductance, plot Fig.~\ref{FigS1}d). It is not simple to determine if those states are topological or not; they could be MBSs when an odd-number of topological subbands are occupied, or trivial ABSs crossing zero, associated with some life time broadening effects (Ref.~\cite{Lee2013}).

\newpage
\paragraph*{\textbf{MBS dot-occupancy dependent measurements}}

\begin{figure}
\renewcommand{\thefigure}{S\arabic{figure}}
\centering \includegraphics[width=16cm]{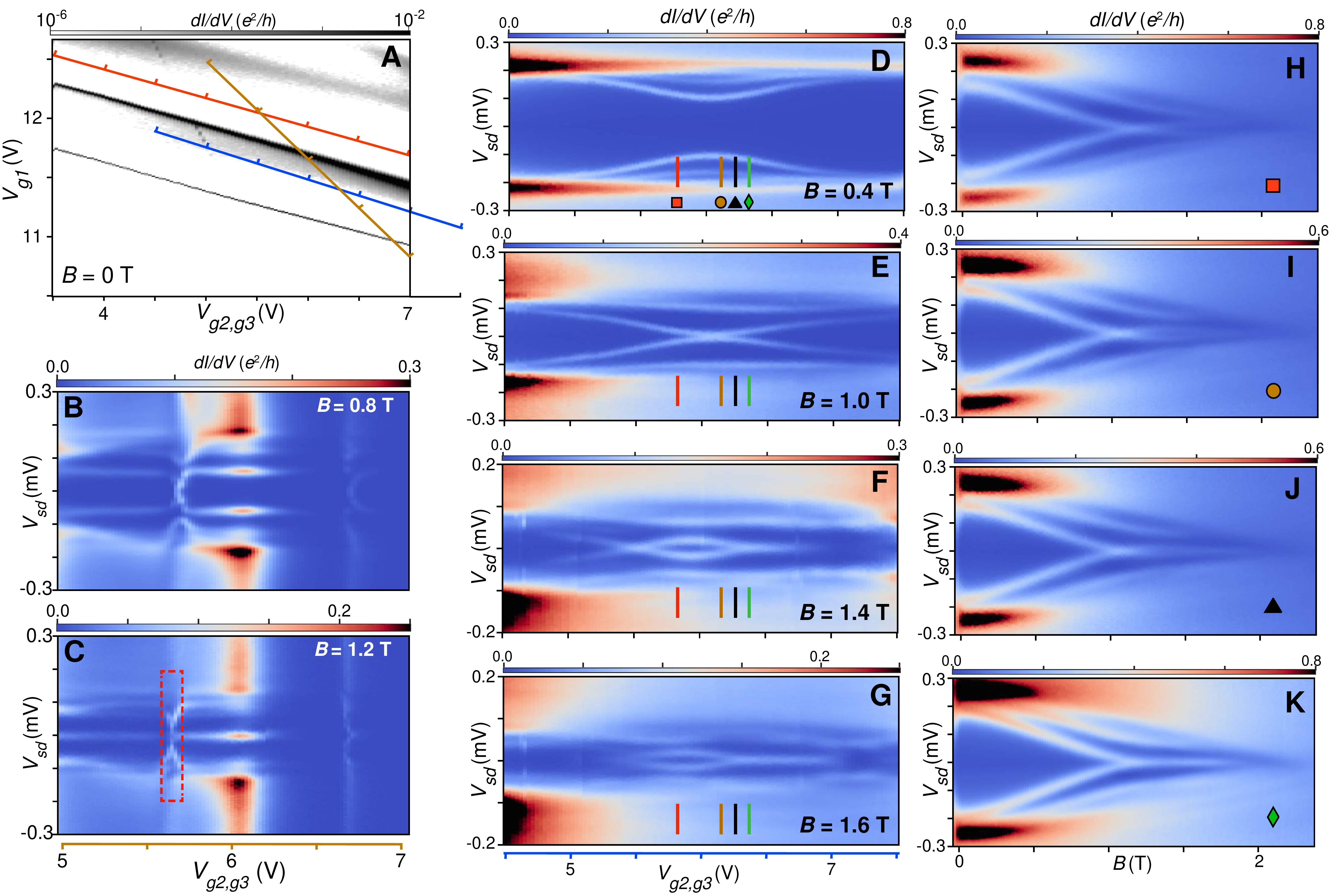}
\caption{\label{FigS2} \textbf{Dot valley dependence of wire subgap states.} \textbf{a}, Charge stability diagram for $V_{g1}$ and $V_{g2,g3}$. It is the same as Fig.~3\textbf{b} of the main article, with the red line representing the sweep direction of the combined gate for Figs.~3\textbf{c-f} of the main article. \textbf{b-c}, The wire-equipotential tunneling spectra, as a function of the combined gate voltage, sweeping along the brown line in \textbf{a}, at $B=0.8$~T and $B=1.2$~T, respectively. They are large scale views of Figs.~4\textbf{c} and \textbf{d} of the main article. The red dashed square region corresponds to the same measuring regime as Fig.~4\textbf{d} of the main article. \textbf{d-g}, Tunneling spectra measured at various magnetic fields, as a function of the combined gate voltage, i.e., along the blue line in \textbf{a}. \textbf{h-k}, $B$-$V_{sd}$ sweeps at different gate voltages, corresponding to the square, triangle, circle and diamond in \textbf{d}, respectively. Depending on gate voltages, the ABSs in the wire show different magnetic field dependence, from a regular ABS behaviour (\textbf{h}) to MBS behaviour (\textbf{k}). }
\end{figure}

In the main article, gate-dependent subgap spectra measured along a dot-equipotential line are shown for an even-occupied dot valley (along the red line in Fig.~\ref{FigS2}a). Here, we show an otherwise identical measurement performed in the adjacent odd-occupied valley (blue line in Fig.~\ref{FigS2}a), as well as for wire-equipotential measurements that cross the dot levels (along the brown line in Fig.~\ref{FigS2}a).

Measurements in the odd-occupied dot valley (Figs.~\ref{FigS2}d-k) show a subgap structure that is similar to the even-valley data in the main article. In Figs.~\ref{FigS2}d-g we show the dependence of the spectrum on the gate voltage (to be compared with Figs.~3c-f in the main article), and in Figs.~\ref{FigS2}h-k the dependence on magnetic field (Figs.~3g-i in the main article). We conclude with showing the spectrum along the wire-equipotential (Figs.~\ref{FigS2}b,c), and see that the subgap levels indeed do not depend on the dot occupancy. Note that the zero-energy state will show an energy splitting when on resonance with the dot, as discussed in the main article. These measurements show that the subgap structure observed in Figs.~3c-i is independent of dot occupancy.

\begin{figure}
\renewcommand{\thefigure}{S\arabic{figure}}
\centering \includegraphics[width=12cm]{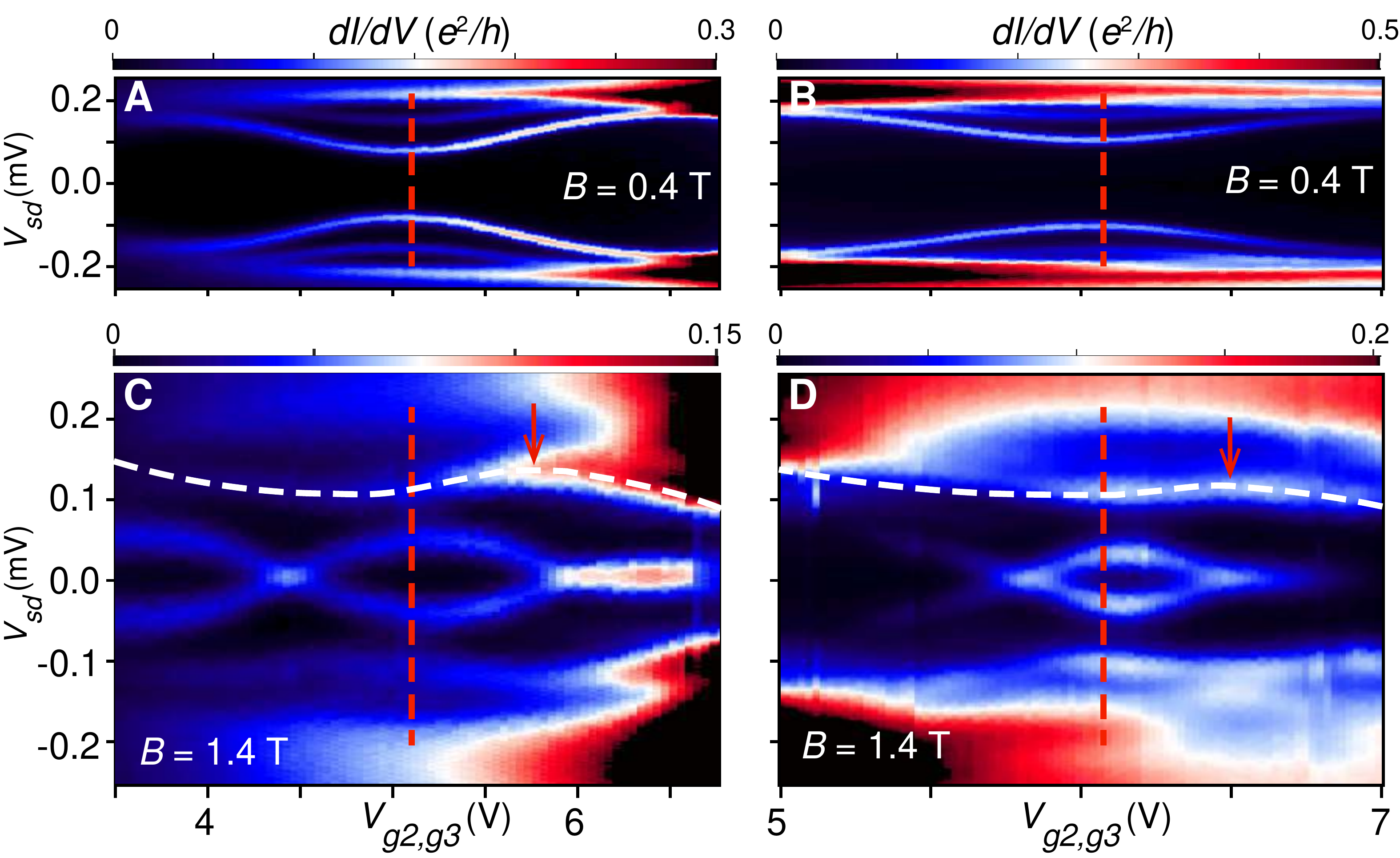}
\caption{\label{FigS3} \textbf{Replotted view of gate dependent subgap spectra.} \textbf{a,c}, Rescaled view of Figs.~3\textbf{d,f} of the main article. The red dashed line is a guide for the eye to show the position of the parabola bottom of the lowest-energy ABS at $B=0.4$~T, and the white dashed line marks the position of the first excited ABS at $B=1.4$~T. The lowest energy ABSs in \textbf{c} are very asymmetric around the red dashed line, with its right part staying at zero-energy and its left part split. Also, note that the excited ABS mode in \textbf{c} appears to originate from two parabolas. The red arrow indicates where the excited state switches from one mode to another. \textbf{b,d}, Similar to \textbf{a,c}, but rescaled versions of Figs.~\ref{FigS2}\textbf{d,f} of the SI. Similar observations as in \textbf{c} are made in \textbf{d}.}
\end{figure}

In Figs.~\ref{FigS3}a,c we replotted on a different color scale for the gate-dependent subgap spectra of Figs.~3d,f in the main article, and in Figs.~\ref{FigS3}b,d we replotted Figs.~\ref{FigS2}d,f of the SI. The lowest energy ABSs in Figs.~\ref{FigS3}c-d are very asymmetric around the position of the parabola bottom at $B=0.4$~T (red dashed lines). Moreover, based on the outline of the excited states (highlighted by the white dashed line), it is clear that the excited states switch from one mode to another around the position of the arrows in Figs.~\ref{FigS3}c-d.

Our interpretation is that the ground state is influenced by a mixing effect between the ground state and different excited states, i.e., whether the ground ABSs split or not depends on the level repulsion between the ground state and the excited state.


\newpage
\paragraph*{\textbf{Longitudinal ABS modes from the same subband}}

\begin{figure}
\renewcommand{\thefigure}{S\arabic{figure}}
\centering \includegraphics[width=12cm]{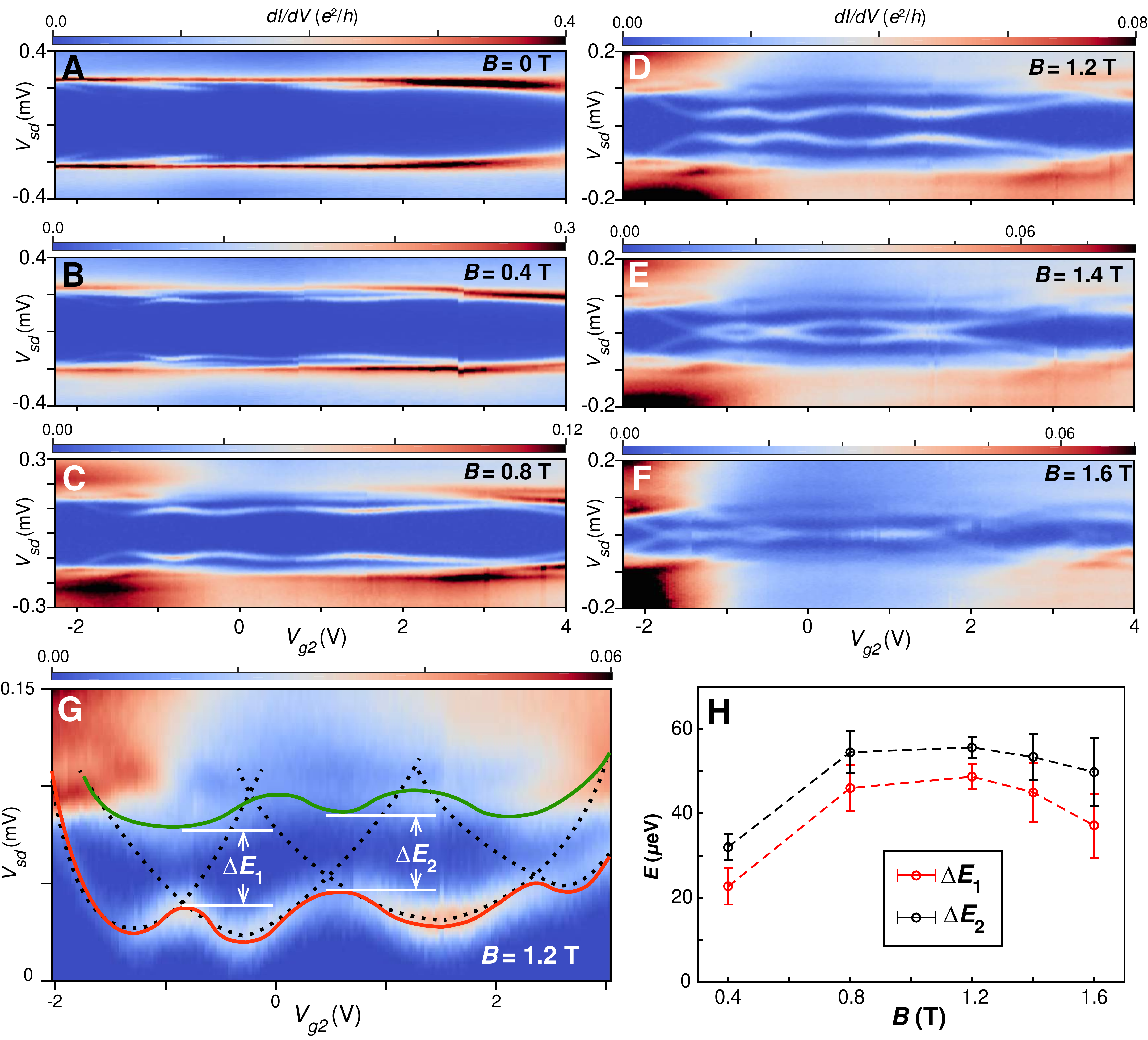}
\caption{\label{FigS4} \textbf{Longitudinal ABS modes from the same subband.} \textbf{a-f}, Tunneling spectra at various magnetic fields, all measured along a dot-equipotential. \textbf{g}, A close-up view of \textbf{d}. Black dashed lines, red solid lines and green solid lines are guides to the eye marking BdG-parabolas without SOI, for the lowest-energy and second-lowest-energy ABSs. All measurements were performed at $V_{bg}=-5.5$~V. \textbf{h}, Extracted anticrossings $\Delta E_{1}$ and $\Delta E_{2}$ in \textbf{g} as a function of magnetic field.}
\end{figure}

In this section we investigate subgap tunneling spectra, showing ABSs which we interpret as modes from the same subband. In Fig.~\ref{FigS4} we show subgap conductance spectra as a function of combined gate voltage (sweeping along a dot-equipotential line; only $V_{g2}$ is shown at the axis), at various magnetic fields.

At zero magnetic field, the induced superconducting gap is hard, with a few weak subgap states close to the bulk (continuous) gap edge. As the magnetic field increases, these subgap states (corresponding to one of the Zeeman-split branches) move towards lower energy. These subgap states show parabola-like structures (Ref.~\cite{Stanescu2013S}), with several parabolas in the range of measured gate voltages. However, there is a large anticrossing wherever two adjacent parabolas cross: We also extracted these two-level anticrossings as a function of magnetic field (Fig.~\ref{FigS4}g).
Level anticrossings are normally caused by state mixing, and SOI-induced state mixing plays a key role for InAs devices (Refs.~\cite{Fasth2007S, Kanai2011S}).

\begin{figure}
\renewcommand{\thefigure}{S\arabic{figure}}
\centering \includegraphics[width=10cm]{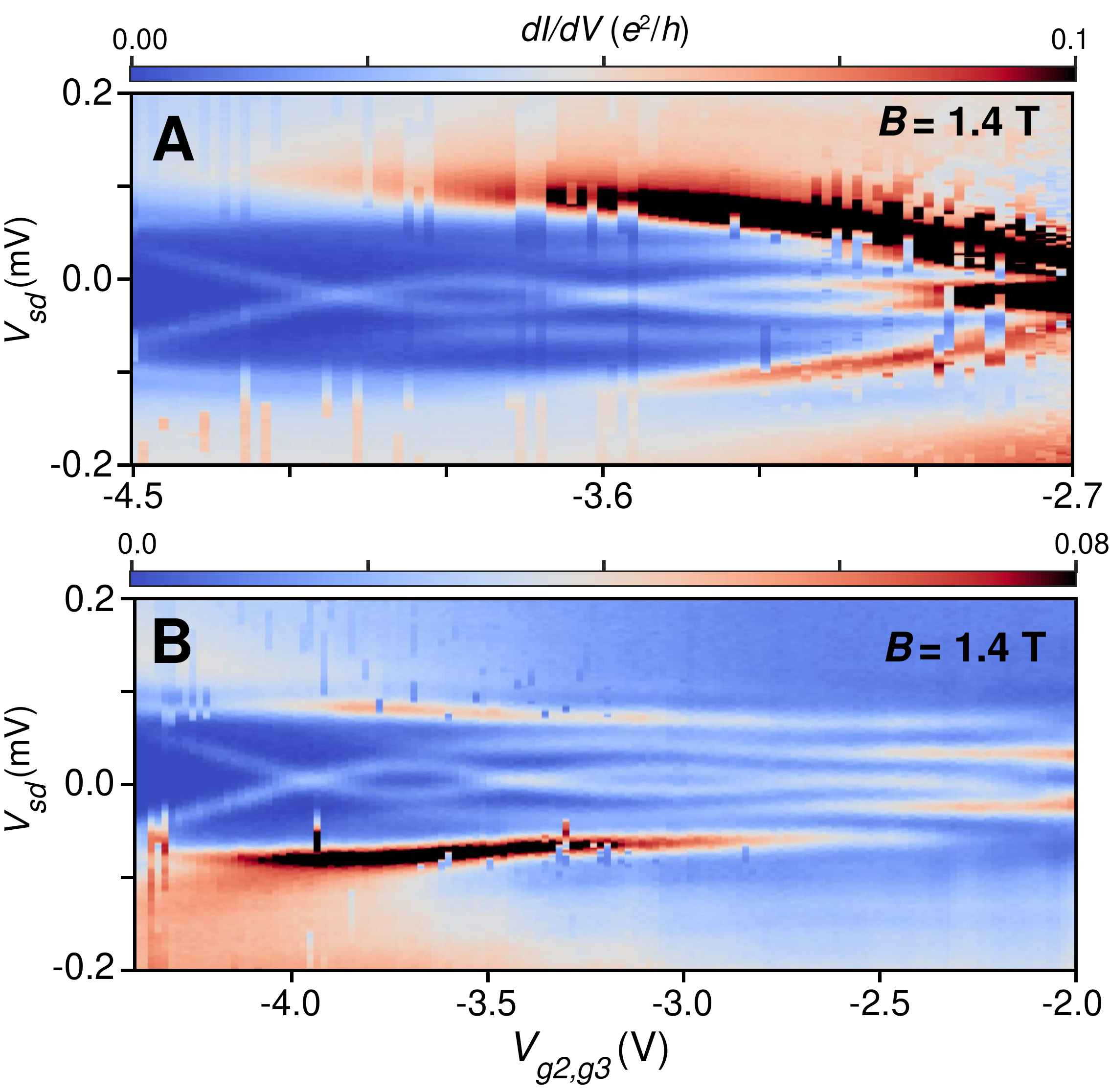}
\caption{\label{FigS5} \textbf{Longitudinal ABS modes from the same subband with disorders.} \textbf{a,b}, Tunneling spectra of device \emph{\textbf{1}} at $V_{bg}=-2.5$~V and $B = 1.4$~T, both are measured along a dot-equipotential. $V_{g1}$ range in \textbf{a} is from 3.2 V to 3.7 V, while in \textbf{b} is from 2.5 V to 3.0 V.}
\end{figure}

The subgap structures in high fields (Figs.~\ref{FigS4}e-f and Figs.~\ref{FigS5}a-b) show clear oscillation patterns, with three wave nodes in each. These oscillatory behaviours are highly consistent with numerical simulation results of single-subband MBSs in the simulation section of this SM and in Refs.~\cite{Rainis2013S, Stanescu2013S}.

MBSs should only appear when odd number of sub-band are occupied. Even number subband occupancy will lead MBSs annihilate at the nanowire ends. However, for a quasi-1D nanowire with a finite length, orbital effects~\cite{Nijholt2015S} and MBS oscillation due to wave function overlapping~\cite{Stanescu2013S} will modify the topological phase diagram a lot.

Technically, smoothly sweeping chemical potential to explore MBSs with different number of sub-bands occupancy at this stage remains challenging, although it is much easier to tune chemical potential comparing to previous superconductor/semiconductor systems. Large scale gate voltage sweeping (especially the back gate sweeping), will dramatically change the constriction conductance/end-dot occupancy. Though the smart gate sweep can compensate the constriction conduction, the ABS/MBS wave function will still be largely stretched. Besides, keeping the chemical potential uniformly (spatially, along the nanowire) tuned is also difficult. For example, Fig.~\ref{FigS5} shows two gate sweeps of MBSs at finite magnetic field, in which the MBSs are interrupted either by non-coalescing ABS (Fig.~\ref{FigS5}a) or by disorders on the constriction (Fig.~\ref{FigS5}b).

\newpage
\paragraph*{\textbf{Zero-bias peaks in strong dot-wire coupling regime}}

\begin{figure}
\renewcommand{\thefigure}{S\arabic{figure}}
\centering \includegraphics[width=15cm]{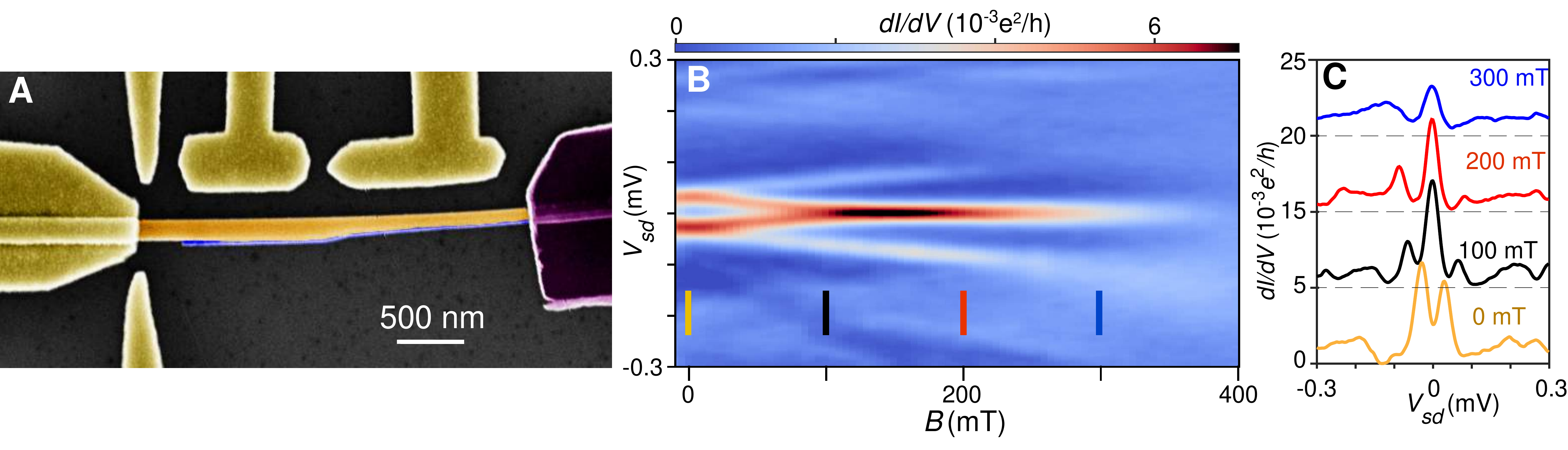}
\caption{\label{FigS6} \textbf{Rigid zero energy state in a strongly coupled dot-wire regime measured on device \emph{\textbf{5}}.} \textbf{a} SEM image of device \emph{\textbf{5}}. \textbf{b} Magnetic field dependence measurements of subgap states in device \emph{\textbf{5}}. A pair of ABS merge into a zero bias peak at $\sim80$~mT and persists to $\sim380$~mT. Due to a relatively thicker Al-shell of this nanowire, the critical magnetic field is much lower than the other devices.
\textbf{c} Line-cuts taken from \textbf{b}.}
\end{figure}

Magnetic field sweep that are similar to Fig.~5 of the main article were also measured for on device \emph{\textbf{5}}, shown in Fig.~\ref{FigS6}.
This measurement was taken in a strongly coupled dot-wire regime, where dot states and wire states can not be distinguished. It shows a zero-bias peak emerging after a pair of inward bound states merges at a finite magnetic field.
The zero-bias peak in Fig.~\ref{FigS6}b is well defined in terms of showing no visible splitting. Comparing to Fig.~6 in the main article, device \emph{\textbf{5}} shows a much lower $B_c$ due to its thicker Al-shell.

\newpage
\paragraph*{\textbf{Magnetic field orientation measurements of subgap states}}

\begin{figure}
\renewcommand{\thefigure}{S\arabic{figure}}
\centering \includegraphics[width=10cm]{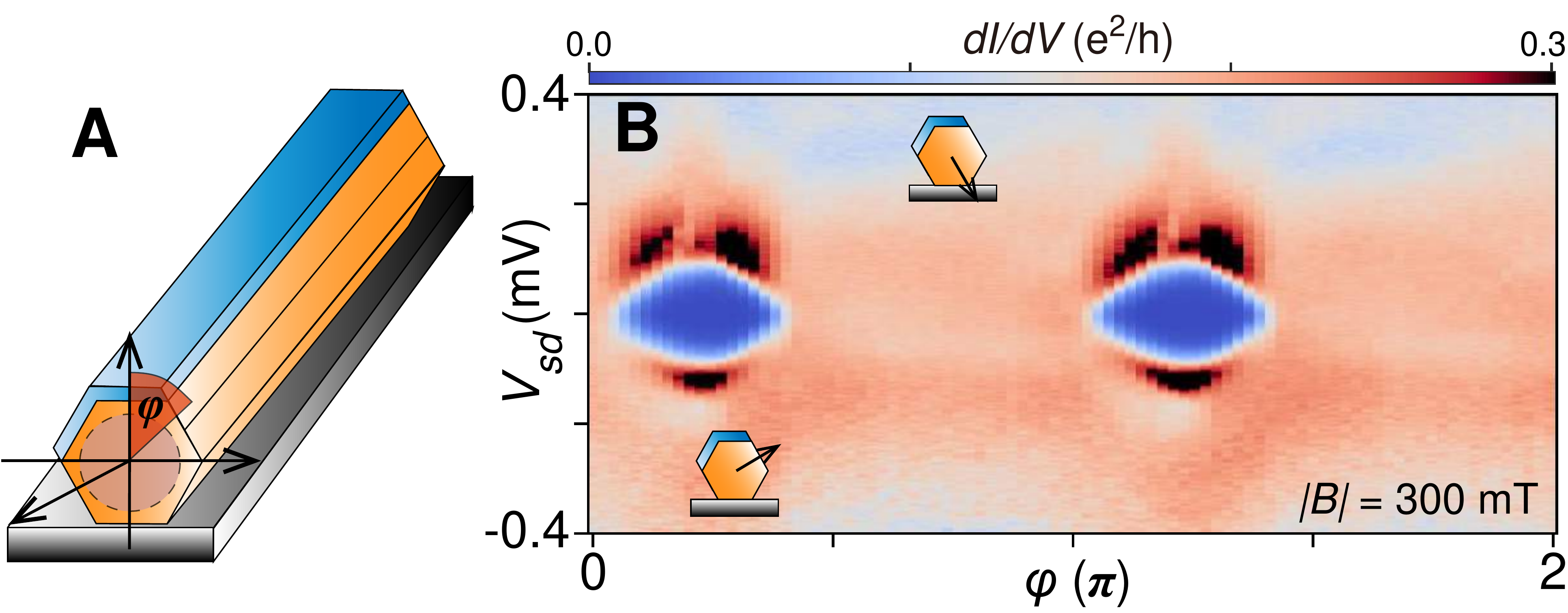}
\caption{\label{FigS7} \textbf{Orientation-dependence of critical field.} \textbf{a}, Schematics of magnetic field rotation directions. The magnetic field rotates in the plane that is perpendicular to the nanowire (when sweeping $\varphi$). \textbf{b}, Differential conductance measured as a function of $\varphi$ and $V_{sd}$, with the magnetic field amplitude fixed at $|B|=300$~mT. The superconducting gap shows a maximum when the magnetic field is roughly in the same plane as the Al shell. The gap gradually vanishes when the magnetic field approaches the direction that is perpendicular to the Al shell.}
\end{figure}

In this section, the dependence of the subgap states on the orientation of the magnetic field is investigated.

First, the applied magnetic field is rotated in the plane perpendicular to the nanowire, that is, around the axis of the wire, with a fixed amplitude (Fig.~\ref{FigS7}).
It is clearly seen that the superconducting gap shows a strong dependence on the magnetic field orientation.
With an amplitude of $|B|=300$~mT, the maximum gap appears when the magnetic field direction is nearly parallel to the two planes of the Al shell (see the diagrams in Fig.~\ref{FigS7}b).
Note that the plane of the Al shell is defined according to the SEM image of the device and material growth information.
The gap quickly vanishes when the magnetic field rotates out of the plane of the Al shell.
This indicates that the critical field is smaller than 300~mT when the angle between the magnetic field direction and the Al shell plane is larger than $\varphi \sim 0.3~\pi$. This critical field is much smaller than the magnetic field at which well defined zero-energy states appear when the field is oriented roughly in the Al plane.

\begin{figure}
\renewcommand{\thefigure}{S\arabic{figure}}
\centering \includegraphics[width=16cm]{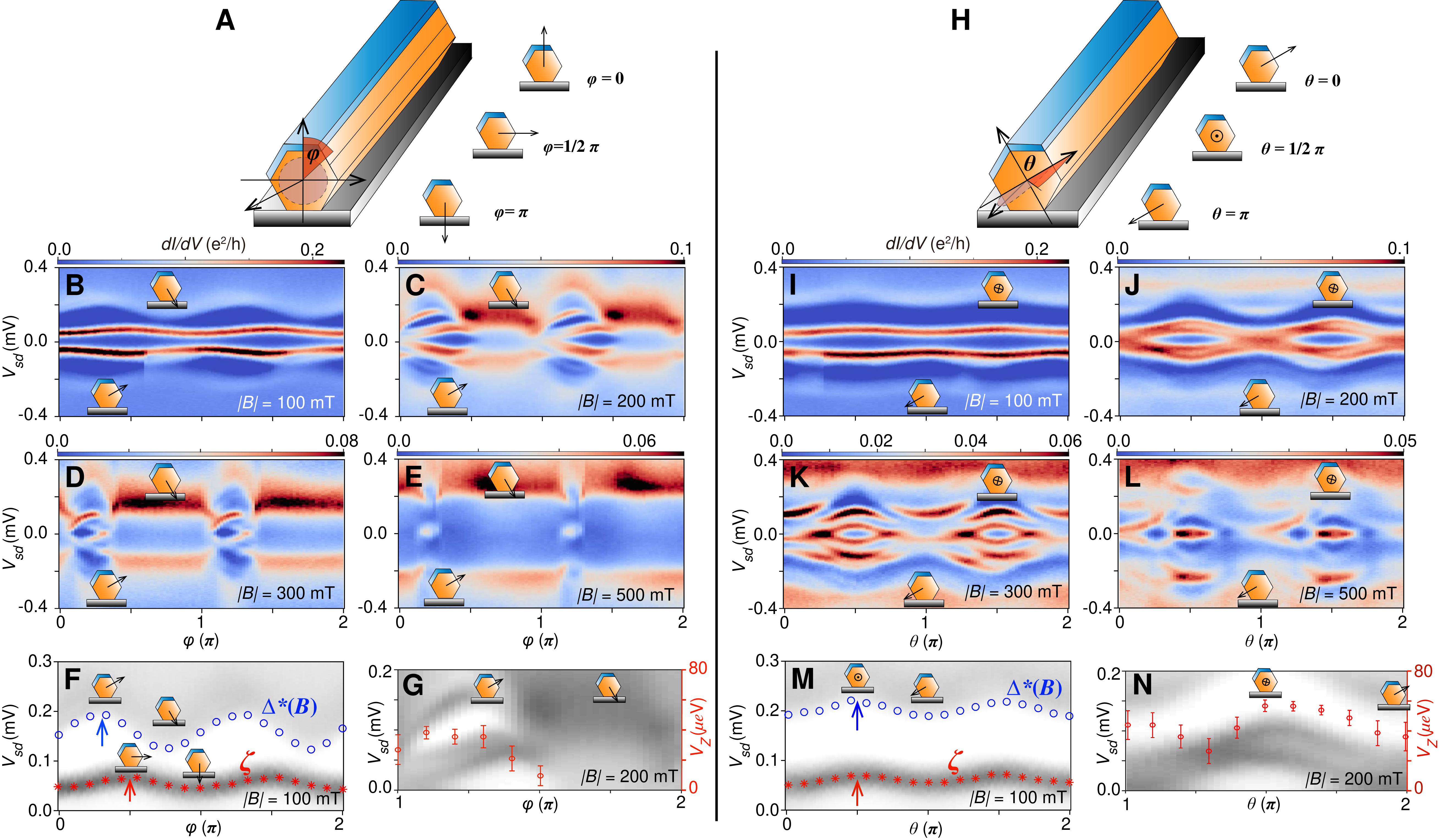}
\caption{\label{FigS8} \textbf{Magnetic field orientation dependence of subgap states.} \textbf{a-g}, The magnetic field is rotated in the plane that is perpendicular to the nanowire (sweeping $\varphi$). \textbf{h-n}, The magnetic field is rotated in the plane that is parallel to the Al shell plane (sweeping $\theta$). \textbf{b-e} and \textbf{i-l}, Differential conductance measured as a function of $\varphi$ or $\theta$, at various fixed magnetic field amplitudes. All measurements are performed at $V_{bg}=-10$~V, $V_{g1}=19$~V, and $V_{g2,g3}=0$. \textbf{f,m}, The superconducting gap $\Delta^*(B)$ (blue circles) and the energy of the subgap state $\zeta$ (red stars) as a function of $\varphi$ or $\theta$, as extracted from \textbf{b} and \textbf{i}, with the corresponding conductance as a greyscale background. The blue and red arrows indicate where the maximum values of the gap and the subgap state energy are. \textbf{g,n}, Zeeman energy splitting as a function of $\varphi$ or $\theta$, extracted from \textbf{c} and \textbf{j}, with the corresponding conductance as a greyscale background. The energy splittings are rescaled to the right axes.}
\end{figure}

Magnetic field orientation dependent measurements of the subgap spectrum are summarized in Fig.~\ref{FigS8}. Besides measurements similar to Fig.~\ref{FigS7}, also data for rotation in the plane of the Al shell (defining by the angle $\theta$) are shown in Figs.~\ref{FigS8}h-n. In Figs.~\ref{FigS8}i-n the value of $\varphi$ is set to the gap maximum point in Fig.~\ref{FigS7}b.

At $|B|=100$~mT, both the effective superconducting gap $\Delta^*(B)$ and the energy of the subgap state $\zeta$ show a sinusoidal variation as a function of $\varphi$ and $\theta$. The $\varphi$-dependence of $\Delta^*(B)$ is consistent with Fig.~\ref{FigS7}, with a maximum appearing when the magnetic field is parallel to the Al shell. The $\theta$-dependence of $\Delta^*(B)$ shows a maximum when the magnetic field is parallel to the wire. This axial field direction, which has the overall largest critical field, is used for the measurements shown in the main article. The subgap states do not show a Zeeman splitting in either the $\varphi$- or the $\theta$-dependent measurements at $|B|=100$~mT. However, the energy of the subgap state $\zeta$ varies in both cases. Interestingly, $\zeta$ oscillates in phase with $\Delta^*(B)$ when $\theta$ is swept, while the oscillations of $\zeta$ and $\Delta^*(B)$ are not in phase when $\varphi$ is swept. The maximum of $\zeta$ as a function of $\varphi$ appears when the magnetic field is parallel to the substrate, and its minimum appears when it is perpendicular to the substrate.

The in-phase oscillations of $\zeta$ and $\Delta^*(B)$ with $\theta$ can be understood as a consequence of the fact that the energy of a given subgap state is correlated with the gap. However, the reason for the phase shift between $\zeta$ and $\Delta^*(B)$ in the $\varphi$-dependent measurements remains unclear. It could be due to orbital effects on the wire states and/or flux focusing/screening effects from the Al shell. It could also be ascribed to the change of the angle between SOI field $B_{SOI}$ and the external field, when considering the $-10$~V back gate voltage could push the electrons in the wire away from the substrate and dominate Rashba asymmetry. This is consistent with the fact that the $\zeta$-$\varphi$ relation is symmetric with respect to the substrate.

A Zeeman splitting of the subgap state is visible at $|B|=200$~mT, in both the $\varphi$- and $\theta$-dependent data. The energy splitting $V_{Z}$ is extracted and plotted in Figs.~\ref{FigS8}g and n. The pronounced variations of $V_{Z}$ indicate a large anisotropy of the effective $g^*$-factor in both cases.

At even higher field, zero-bias peaks start to show up for some field orientations. However, it is difficult to track which zero-bias peaks are a signature of a MBS and which are just due to zero-crossings of regular ABSs, due to the anisotropies of $B_c$, $\zeta$, the $g^*$-factor and SOI.

\newpage
\paragraph*{\textbf{Simulation of Majorana-dot interaction}}

\vspace{20pt}
\begin{figure}
\renewcommand{\thefigure}{S\arabic{figure}}
\centering \includegraphics[width=8cm]{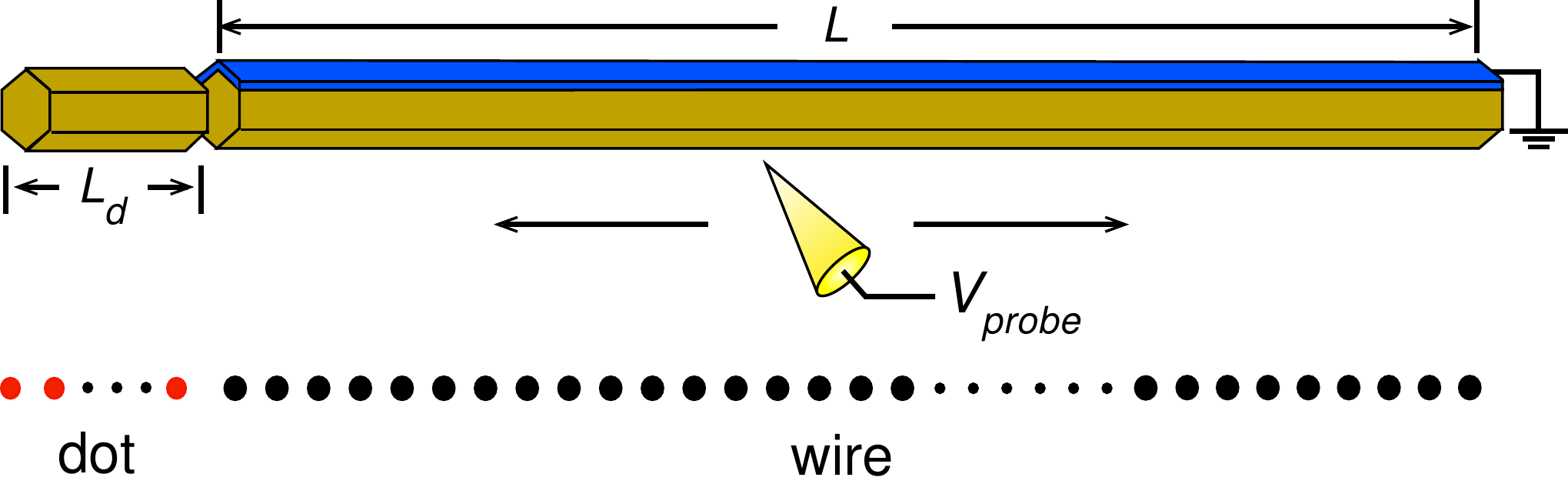}
\caption{\label{FigS9} \textbf{Schematic of the model used in this work.} The nanowire is partly covered by a bulk superconductor (indicated in blue). Immediately left of the superconducting region of length $L$ there is a potential barrier in the wire that effectively creates a quantum dot of length $L_d$ at the left end of the wire. A tunnel probe can be moved along the wire to measure the local differential conductance.}
\end{figure}

Our simulations consist of numerical calculations of the differential conductance of the wire as measured through a movable tunnel probe, see the sketch in Fig.~\ref{FigS9}.
The right part of the effectively one-dimensional wire is proximity-coupled to a grounded bulk $s$-wave superconductor.
The resulting proximity-induced superconducting region of length $L$ is connected through a potential barrier of width $L_b$ to a normal island (quantum dot) of width $L_d$ at the left end of the wire.
The tunnel probe measures the local differential conductance of the wire at position $z_{\rm probe}$, with $0 < z_{\rm probe} < (L_d +L_b + L)$.

\begin{figure}
\renewcommand{\thefigure}{S\arabic{figure}}
\centering \includegraphics[width=14cm]{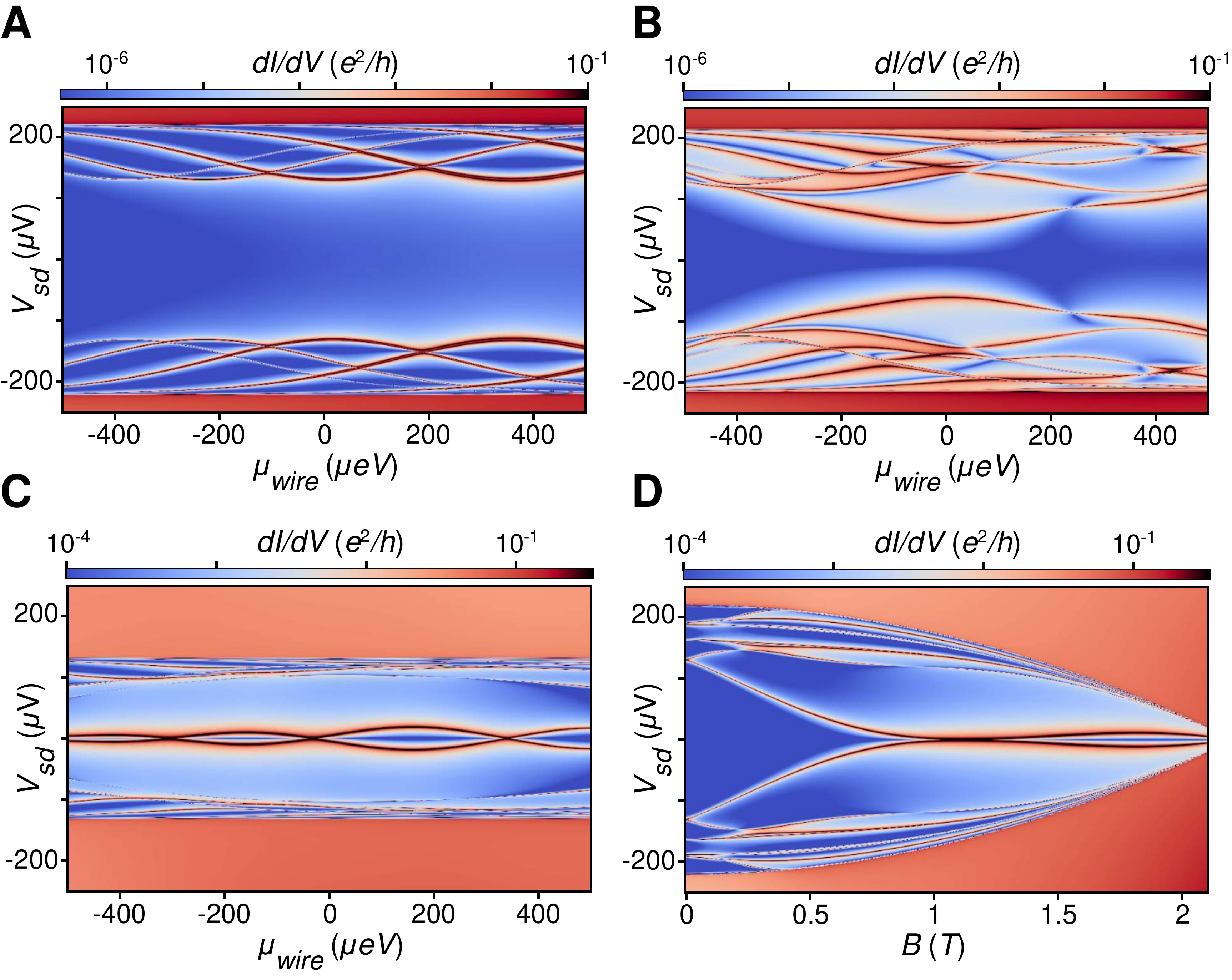}
\caption{\label{FigS10} \textbf{Numerical simulation of the dot-wire system.} Calculated differential conductance spectrum as measured through the tunnel probe with $z_{\rm probe}=0$, to be compared with Figs.~3c-g in the main article. All numerical parameters are specified in the text. \textbf{a-c} Conductance as a function of $\mu_{\rm wire}$ for $B=0$, $0.4$, and $1.4$~T, respectively. \textbf{d} Magnetic field dependence of the conductance spectrum at $\mu_{\rm wire}=0$.}
\end{figure}

 \vspace{20pt}
Using a one-channel model, we start from a one-dimensional Bogoliubov-de Gennes Hamiltonian for the electrons and holes in the nanowire (see e.g. \cite{Hansen2015S} and references therein),
\begin{equation}
  \mathcal{H}_{\text{NW}} = \left[-\frac{\hbar^2\p_z^2}{2m^*}-i\alpha\p_z\sigma_y + V(z)\right]\tau_z + E_{\rm Z}\sigma_z,
\end{equation}
where $m^*$ is the effective mass of the electrons in the wire, $\alpha$ is the strength of the spin-orbit coupling, $E_{\rm Z}=\tfrac{1}{2} g \mu_{\rm B} B$ is the Zeeman energy, and
\begin{equation}
V(z) =
\begin{cases}
-\mu_{\rm dot} + \tfrac{1}{2}E_C(\mathds{1}_2+ \sigma_z\tau_z)& \text{for } z < L_d, \\
V_0 & \text{for } L_d \leq z \leq L_d +L_b, \\
-\mu_{\rm wire} & \text{elsewhere,}
\end{cases}
\label{eq:vz}
\end{equation}
so that $\mu_{\rm dot}$ is the chemical potential on the dot, $V_0$ characterizes the height of the barrier, and $\mu_{\rm wire}$ is the chemical potential on the proximitized part of the wire.
The Pauli matrices $\boldsymbol \sigma$ and $\boldsymbol \tau$ act in spin space and electron-hole space respectively.
Charging effects on the normal island are accounted for phenomenologically by the term $E_C$ in $V(z)$, which lifts the Kramers degeneracy in the dot states at $E_Z = 0$ and qualitatively reproduces the results of a mean-field description  of the Coulomb interactions at finite magnetic field.

For our numerical simulations, we use a tight-binding approximation to discretize this Hamiltonian on a lattice with $N$ sites, where we used $N=100$ and $N=200$ for different plots.
The Green function (matrix) for the electrons and holes in the wire then follows as $G^R(\epsilon)=\left[\epsilon-{\cal H}_{\text{NW}}-\Sigma+i0^+\right]^{-1}$, where the self energy
\begin{equation}\label{selfenergy}
\Sigma(z,z';\epsilon) = \gamma(z) \frac{-\epsilon + \Delta\tau_x}{\sqrt{\Delta^2-(\epsilon+i0^+)^2}}\delta_{z,z'},
{}
\end{equation}
results from integrating out the degrees of freedom of the bulk superconductor.
Here, $\Delta$ is the superconductor's pairing potential, $0^+$ is a positive infinitesimal, and
\begin{equation}
\gamma(z) =
\begin{cases}
\gamma & \text{for }L_d+L_b < z < L_d +L_b + L, \\
0 & \text{elsewhere,}
\end{cases}
\end{equation}
parametrizes the coupling between the wire and the superconductor.

From the Green function we can find the scattering (reflection) matrix of the wire as
\begin{align}
R(\epsilon) &= \twomatrix{r_{\text{ee}}(\epsilon)}{r_{\text{eh}}(\epsilon)}{r_{\text{he}}(\epsilon)}{r_{\text{hh}}(\epsilon)}\nonumber\\
&=1-2i\pi W^\dag\left\{ \left[G^R(\epsilon)\right]^{-1}+i\pi WW^\dag\right\}^{-1}W,
\end{align}
where $r_{\rm ee(hh)}$ are the normal electron(hole) reflection amplitudes and $r_{\rm eh,he}$ describe Andreev reflection.
The coupling to the tunnel probe is included through the matrix
\begin{equation}
W = \sqrt{\gamma_W}\left(\mathbf{s}_n\otimes \mathds{1}_4\ri)^{\text{T}}
\end{equation}
where $\gamma_W$ parametrizes the coupling strength. The vector $\mathbf{s}_n = (0,0,...,0,1,0,...)$ has a single $1$ located at the $n$'th entry and thereby encodes the position of the tunnel probe, $z_{\rm probe} = n(L_d+L_b+L)/N$.

The resulting reflection matrix $R(\epsilon)$ then allows us to
calculate the differential conductance as
\begin{equation}
\frac{dI}{dV} = \frac{e^2}{h}\text{Tr}\left[1-|r_{\text{ee}}(eV_{\rm bias})|^2+|r_{\text{eh}}(eV_{\rm bias})|^2\right],
\end{equation}
where $V_{\rm bias}$ is the bias voltage of the tunnel probe. More details about this calculation can be found in Ref.~\cite{Hansen2015S}.

\vspace{20pt}
In Fig.~\ref{FigS10} we show the calculated differential conductance probed from the left end of the wire ($z_{\rm probe}=0$) as a function of source-drain bias voltage (corresponding to $\epsilon$) and $\mu_{\rm wire}$ (in Figs.~\ref{FigS10}a-c) and applied magnetic field $B$ (in Fig.~\ref{FigS10}d).
The plots shown in Figs.~\ref{FigS10}a-c have $B=0$, $0.4$, and $1.4$~T, respectively, and in Fig.~\ref{FigS10}d we set $\mu_{\rm wire}=0$.
In all these plots we used $m^* = 0.026\,m_e$, $\alpha = 0.5$~eV\AA, $g=-18$, $\mu_{\rm dot} = 4$~meV, $E_C = 10$~meV, $V_0 = 8$~meV, $L_d = 60$~nm, $L_b=12$~nm, $L=528$~nm, $\gamma = 256~\mu$eV, and $\gamma_W = 2.2$~meV.
For this value of $\mu_{\rm dot}$ there are no dot levels close to zero energy, and the system is thus tuned to the ``cotunneling'' regime.
The parameter $\Delta$ was made $B$-dependent, $\Delta(B) = \Delta(0) [1 - (B/B_c)^2]$ to phenomenologically describe the aluminum gap closing at higher fields, which reproduces the experimental gap closing seen to Figs.~3g-i of the main article.
Based on the data presented in Figs.~3g-i of the main article, we set $\Delta(0) = 220~\mu$eV and $B_c = 2.2$~T.
We note that the values used for $\alpha$ and $\gamma$ were also extracted from the experimental data (see next section).
The rather large $g$-factor we used has to be seen as a phenomenological parameter and not as the ``bare'' wire $g$-factor, since our model does not include $g$-factor renormalization by the strong coupling to the aluminum shell.
It is also hard to fit the effective $g$-factor unambiguously from the data since finite-size effects make it very hard to read off accurately the critical field of the topological phase transition.
We however see that our simple model reproduces qualitatively the $\mu_{\rm wire}$-dependence of the ABSs close to $\mu_{\rm wire} = 0$, including the characteristic ``low-energy oscillations'' at higher fields.

In Figs.~4e,f of the main article we show the conductance spectrum as a function of $\mu_{\rm dot}$ and $\epsilon$, with $\mu_{\rm wire}$ set to zero.
We tune $\mu_{\rm dot}$ such that there now exist a dot level close to zero energy.
For both these figures we used the same parameters as above, but adjusted the coupling parameter $\gamma_W = 147~\mu$eV.
In Fig.~4e we used $B=0.4$~T and in Fig.~4f $B=1.04$~T.
We see that our numerics again qualitatively capture all observed phenomena: At low fields (in the trivial regime) the dot level crosses zero energy and anticrosses with all ABSs. At higher fields (after the gap closing), the anticrossing of the dot level with the low-energy MBSs results in a characteristic ``eye'' feature in the spectrum.

In the main article, we interpreted this eye feature as resulting from a hybridization between the low-energy MBSs and the dot level: The MBSs partly leak onto the dot, which causes the effective overlap of the two end-state wave functions to change. Since in a finite wire the actual splitting between the two MBSs depends sensitively on the details of their wave function, this splitting can significantly change on resonance. In case the splitting is vanishingly small far from resonance, it can can increase when approaching the resonance and have a maximum when the dot level is at zero energy.

\vspace{20pt}
\begin{figure}
\renewcommand{\thefigure}{S\arabic{figure}}
\centering \includegraphics[width=14cm]{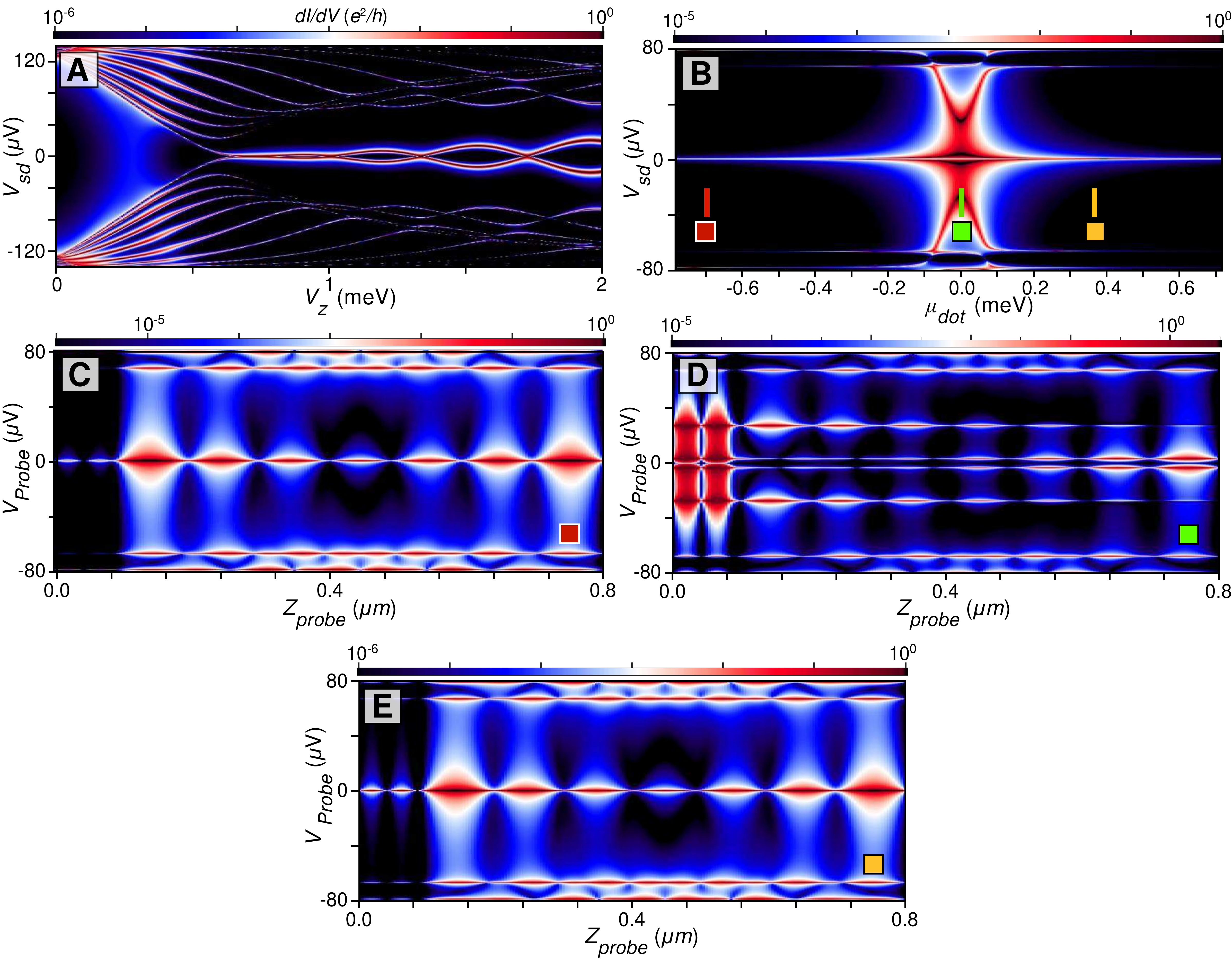}
\caption{\label{FigS11} \textbf{Numerical investigation of the resonant dot-wire coupling.} \textbf{a}, Simulated subgap state tunneling spectrum with the movable probe at the left end of the system ($z_{\rm probe}=0$), as a function of Zeeman energy $V_{\rm Z}$. After the gap closing, the energy of the finite-sized MBSs oscillates as the Zeeman energy increases further. \textbf{b}, Differential conductance as a function of the dot chemical potential $\mu_{\rm dot}$ and source-drain bias voltage in the topological regime, where a dot level is swept through zero energy. \textbf{c-e}, Differential conductance as a function of bias voltage and position of the tunneling probe $z_{\rm probe}$,  for various dot potentials $\mu_{\rm dot}$, as indicated in the plot. Note that for these plots we used parameters slightly different from those used for the other figures.}\end{figure}

In Fig.~\ref{FigS11} we investigate this effect in more detail.
The figure presents again numerical calculations of the differential conductance measured through the tunnel probe, but to emphasize the physics we used slightly different parameters:
We changed $\alpha = 0.2$~eV\AA, $V_0 = 30$~meV, $L_d = 80$~nm, $L_b=16$~nm, $L=704$~nm, $\gamma = 560~\mu$eV, $\gamma_W = 100~\mu$eV, and $\Delta(B) = 140~\mu$eV.
In Fig.~\ref{FigS11}a we see a typical Majorana-wire conductance spectrum: The gap closes at $V_{\rm Z} \sim 0.6$~meV, after which there is a clearly visible low-energy mode that splits and oscillates at higher Zeeman fields.
Fig.~\ref{FigS11}b has $V_{\rm Z}$ tuned to the topological regime and shows the conductance spectrum when a dot level is swept through zero energy, cf.~Figs~4d,f in the main article.
To support our interpretation of the eye feature as being due to leakage of the MBS wave function onto the dot, we show in Figs.~\ref{FigS11}c-e the differential conductance as a function of the position of the tunnel probe, thus mapping out a local tunneling DOS.
Figs.~\ref{FigS11}c-e correspond to three different dot level tunings (as indicated in Fig.~\ref{FigS11}b with the colored squares), corresponding respectively to far below resonance, at resonance, and a bit above resonance.
We see that close to the resonance the low-energy modes indeed shift part of their weight onto the dot sites, which is accompanied by an increased splitting between the modes.

\newpage
\paragraph*{\textbf{Extracting parameters from the observed spectrum}}

In this section, we discuss how to relate the observed conductance spectrum to an estimate for the spin-orbit interaction in the wire, following the paper by van Heck et al.\cite{Heck2016S}. A similar analysis of finite size effects (using a frequency independent proximity induced pairing) was done by Mishmash et al.\cite{Mishmash2016S}.

The observed parameters are the gap in the superconducting (aluminum) shell $\Delta$, the induced gap in the wire at zero field $\Delta_{ind}$, and the length of the wire $L$.
The analysis starts from the Green function for electrons in the wire, which have a self energy due to the tunneling coupling to the superconducting shell,
\begin{equation}\label{Gm1}
  G^{-1}(k,\omega)=\omega - H_\mathrm{NW}-
\gamma \frac{ \Delta\tau_x-\omega}{\sqrt{\Delta^2-\omega^2}}+i0^+,
\end{equation}
see also Eq.~\eqref{selfenergy}.
The Rashba wire Hamiltonian is in $k$-space given by
\begin{equation}\label{HNWk}
  H_\mathrm{NW} =\left(\frac{\hbar^2k^2}{2m^*}-\mu+\alpha k\sigma_y \right)\tau_z + \frac12 g\mu_B B\sigma_z.
\end{equation}
For energies below the gap $E<\Delta$ the $k$-dependent electronic spectrum described above follows from solving
\begin{equation}
{\rm det} \big[  {\rm Re}\, G^{-1}(k,E) \big]=0.
\end{equation}
We can get an estimate for $\gamma$ by relating this equation to the zero-field observed induced gap.
Setting $B=0$, $\mu=0$, and $k=0$, the lowest available energy follows from setting $E=\Delta_{ind}$, yielding
\begin{equation}\label{Gm1}
  \Delta_{ind}\left(1+\frac{\gamma}{\sqrt{\Delta^2-(\Delta_{ind})^{2}}}\right) -\frac{\gamma\Delta}{\sqrt{\Delta^2-(\Delta_{ind})^{2}}} =0.
\end{equation}
We see that this allows us to express $\gamma$ in terms of $\Delta$ and $\Delta_{ind}$ as
\begin{equation}\label{gammadef}
  \gamma=\Delta_{ind}\sqrt{\frac{\Delta+\Delta_{ind}}{\Delta-\Delta_{ind}}}.
\end{equation}
Using the experimental parameters $\Delta = 220~\mu$eV and $\Delta_{ind} = 130~\mu$eV (determined from the data shown in Fig.~3c) we thus find $\gamma \approx 256~\mu$eV.

From investigating the low-energy spectrum at the topological phase transition in a similar way, we can also arrive at an estimate for $\alpha$, the strength of the spin-orbit interaction.
The spectrum for $\omega\ll \Delta$ near the topological transition can be found from solving (valid for $\mu=0$ and small $k$)
\begin{equation}\label{Gm1}
{\rm det} \big[ {\rm Re}\, G^{-1}(k,\omega) \big]
\approx {\rm det}\left[ \omega\left(1+\frac{\gamma}{\Delta^\prime}\right) -\alpha k\sigma_y \tau_z - \frac12 g\mu_B B\sigma_z -\gamma \tau_x \right] =0,
\end{equation}
where $\Delta^\prime$ is the gap of the aluminum shell at the transition point.
Close to this point we thus have
\begin{equation}
E_{k}\approx \frac{\Delta^\prime}{\Delta^\prime+\gamma}\sqrt{\left(\tfrac12 g\mu_BB-\gamma\right)^{2}+k^{2}\alpha ^{2}}\approx \frac{\Delta^\prime\alpha|k| }{\Delta^\prime+\gamma},\quad \text{at} \quad B=B_{\mathrm{c,topo}},
\label{Ek}
\end{equation}
where $B_{\mathrm{c,topo}}$ is the field at which the wire enters the topological regime.
For an infinite wire, $B_{\rm c,topo}$ would be at the point where the gap closes.
For a finite wire the phase transition happens at the same field as in the infinite wire but the gap closing point moves to higher field, making it difficult to read off a precise value for $B_{\rm c,topo}$.
From \eqref{Ek}, we get that at $B_{\rm c,topo}$ the gap to the first excited state is
\begin{equation}
\delta \approx \frac{\alpha R\pi  }{L}
\quad \mathrm{or\quad }
\frac{\alpha}{L}
\approx \frac{\delta}{R\pi},\quad \mathrm{with\quad }R=\frac{\Delta^\prime}{\Delta^\prime+\gamma}.
\end{equation}
If we use $\xi\approx\alpha/\Delta^\prime$ as the estimation of the coherence length near the phase transition, then we have
\begin{equation}
L/\xi\approx R\pi \Delta^\prime/\delta
\end{equation}
We estimate from Fig.~3h that $B_{\rm c,topo} \approx 1000$~mT, and for those field values we read off $\delta \approx 100~\mu$eV and $\Delta^\prime \approx 180~\mu$eV, yielding
\begin{equation}
L/\xi\approx2.3
\end{equation}
\vspace{20pt}

%

\end{document}